\documentclass[10pt, conference, letterpaper]{IEEEtran}

\usepackage{cite}

\ifCLASSINFOpdf
   \usepackage[pdftex]{graphicx}
   \DeclareGraphicsExtensions{.pdf,.jpeg,.png}
\else
   \usepackage[dvips]{graphicx}
   \DeclareGraphicsExtensions{.eps}
\fi
\usepackage{epstopdf}

\usepackage{amsmath}
\usepackage{mathtools}
\usepackage{amssymb}

\usepackage{array}

\ifCLASSOPTIONcompsoc
  \usepackage[caption=false,font=normalsize,labelfont=sf,textfont=sf]{subfig}
\else
  \usepackage[caption=false,font=footnotesize]{subfig}
\fi

\usepackage{url}

\usepackage{algorithm}
\usepackage{algpseudocode}
\usepackage{mdframed}
\usepackage{varwidth}
\usepackage{flushend}

\begin{document}
        \newtheorem{theorem}{Theorem}
        \newtheorem{proposition}{Proposition}
        \newtheorem{lemma}{Lemma}
        \newtheorem{corollary}{Corollary}
        \newtheorem{definition}{Definition}
        \newtheorem{problem}{Problem}
        \newcommand{\ud}{\,\mathrm{d}}
        \DeclarePairedDelimiter\pparen{\lparen}{\rparen}
        \DeclarePairedDelimiter\pbrace{\lbrace}{\rbrace}

        \makeatletter
        \newlength{\trianglerightwidth}
        \settowidth{\trianglerightwidth}{$\triangleright$~}
        \algnewcommand{\LineCommentCont}[1]{\Statex \hskip\ALG@thistlm%
          \parbox[t]{\dimexpr\linewidth-\ALG@thistlm}
        {\leftskip=\algorithmicindent
          \hangindent=\algorithmicindent
          \hangafter=1%
          \strut\makebox[\algorithmicindent][c]{$\triangleright$}#1\strut}
          } 
        \algnewcommand{\MyState}[1]{\State
        \parbox[t]{\dimexpr\linewidth-\ALG@thistlm}{\hangindent=\algorithmicindent\strut\hangafter=1#1\strut}}
        \algnewcommand{\MyStatex}[1]{\Statex
        \parbox[t]{\dimexpr\linewidth-\ALG@thistlm}{\hangindent=\algorithmicindent\strut\hangafter=1#1\strut}}        \algnewcommand{\Or}{\textbf{or}}
        \algnewcommand{\And}{\textbf{and}}
        \makeatother

        \setlength{\floatsep}{-10pt}


\title{Online Load Balancing for Network Functions Virtualization}

\author{
\IEEEauthorblockN{Tuan-Minh Pham\IEEEauthorrefmark{1},
        Thi-Thuy-Lien Nguyen\IEEEauthorrefmark{1},
        Serge Fdida\IEEEauthorrefmark{2},
    Huynh Thi Thanh Binh\IEEEauthorrefmark{3}
        }
\IEEEauthorblockA{\IEEEauthorrefmark{1}Hanoi National University of Education, Vietnam}
\IEEEauthorblockA{\IEEEauthorrefmark{2}UPMC Sorbonne Universit\'es, France\\
\IEEEauthorblockA{\IEEEauthorrefmark{3}Hanoi University of Science and Technology, Vietnam\\
Email: \{minhpt, lienntt\}@hnue.edu.vn, serge.fdida@lip6.fr, binhht@soict.hust.edu.vn}}
}

\maketitle

\begin{abstract}
        Network Functions Virtualization (NFV) aims to support service providers to deploy various services in a more agile and cost-effective way.
                However, the softwarization and cloudification of network functions can result in severe congestion and low network performance.
                In this paper, we propose a solution to address this issue. We analyze and solve the online load balancing problem using multipath routing in NFV to optimize network performance in response to the dynamic changes of user demands.
                In particular, we first formulate the optimization problem of load balancing as a mixed integer linear program for achieving the optimal solution.
                We then develop the ORBIT algorithm that solves the online load balancing problem. The performance guarantee of ORBIT is analytically proved in comparison with the optimal offline solution.
                The experiment results on real-world datasets show that ORBIT performs very well for distributing traffic of each service demand across multipaths without knowledge of future demands, especially under high-load conditions.
\end{abstract}


%
\IEEEpeerreviewmaketitle

\section{Introduction}
        Communication networks have been growing rapidly with an increased growth of network-based services over the past decades.
                In such a context, network service providers have to reduce their operational costs and time-to-market for network services in order to adapt their business to on-demand  customer needs.
                Responding to these objectives, Network Functions Virtualization (NFV) is a recent trend of network transformation that helps service providers offer new and multiple services in a more agile and cost-effective way.
                By building a virtualized infrastructure where network functions (e.g., Network Address Translation (NAT), Deep Packet Inspection, firewall) are softwarized and virtualized instead of embedded in specialized hardware devices (i.e., middlebox), NFV has the potential to revolutionize the entire telecommunication industry \cite{nfvrg_irtf,etsi2014_architectural}.

        Despite the potential of NFV, critical considerations in traffic engineering must be taken into account to maintain strict performance requirements of virtual network functions (VNF), alike in traditional networks \cite{mijumbi2016_network}.
                In this context we tackle the value of using load balancing (LB) to support NFV deployment. Indeed, LB is an important approach to traffic engineering in the Internet as it  splits the traffic among multiple paths in order to optimize link utilization, reduce congestion, and minimize delay of data flows in the network.
                However, to the best of our knowledge, there has been no attempt thus far to propose an effective solution for online load balancing using multipath routing in NFV.

        A widely used load balancing technique is ECMP (Equal-cost Multipath).
                The principle of ECMP is that the total traffic going out from one network node will be divided equally over all the shortest paths to the same destination with the same cost.
                Many routing protocols like Open Shortest Path First (OSPF) and Intermediate System to Intermediate System (IS-IS) support ECMP \cite{moy1998_ospf,callon1990_isis}.
                Constraints on the shortest path routing and equal load sharing makes the problem complex.
                It becomes even more difficult when considering the NFV characteristics such as the service function chaining (SFC) and the virtualization capacity at nodes and links of the NFV infrastructure (NFVI).

        In this paper, we address an optimization problem of load balancing for NFV, which takes into account the fundamental features of NFV, constraints on the NFVI resource, and ECMP routing.
                The first main contribution of the paper is an analysis and modeling of the load balancing problem using ECMP for NFV.
                We formulate the optimization problem of load balancing across multiple paths in NFV as a Mixed Integer Linear Programming (MILP) model.
                The second important contribution is the ORBIT algorithm that provides an efficient solution for the online load balancing problem.
                The performance guarantee of ORBIT is analytically proved in comparison with the optimal offline solution.

        We evaluated our load balancing solution using two real-world datasets.
                The results show that ORBIT can provide a solution that is close to optimal.
                Specially, our algorithm performs more efficiently in a large-scale NFVI under high traffic load conditions.
                Our solution can be deployed in a SDN network controller where the forwarding rules can be distributed to the forwarding plane of a network switch by using OpenFlow \cite{mckeown2008_openflow}.

        The rest of the paper is organized as follows.
                Section \ref{s-related} reviews the related work.
                Section \ref{s-problem} describes the load balancing problem across multiple paths and formulates the optimization problem as a MILP model.
                Section \ref{s-online} describes our proposed algorithm for solving the online version of the problem and our theoretical analysis of the algorithm performance.
                Section \ref{s-evaluation} presents the performance evaluation of our solution.
                Section \ref{s-conclusion} concludes the paper and highlights future work.

\section{Related Work}
        \label{s-related}
        Many problems arising in the implementation of network functions virtualization have been considered such as the performance and architecture of NFV, and VNF management.
                The research team of Internet Research Task Force (IRTF) and European Telecommunications Standards Institute (ETSI) are both developing standards for NFV \cite{nfvrg_irtf,etsi2014_architectural}.
                Some research results about NFV are summarized in \cite{mijumbi2016_network,xia2015_optical}.
                More recent results about the performance of NFV are mentioned in \cite{elias2014_optimization,cohen2015_near,mohammed2016_sdn,zhang2015_routing,pham2016_load,leivadeas2016_dynamic}.
                There are many studies about load balancing for IP network in the current Internet architecture. \cite{fortz2000_internet,burns2003_path,raiciu2011_improving,patel2013_ananta}.
                However, so far, no research has been found that surveyed an optimal model and efficient online algorithms for load balancing across multiple paths for NFV.

        Several papers were published to study the traffic engineering problem in NFV for better performance and efficient resource utilization \cite{mohammed2016_sdn,zhang2015_routing,pham2016_load,leivadeas2016_dynamic}.
                In \cite{mohammed2016_sdn}, the authors provide the SDN controller's design for load-balanced network resources usage. In their design, if a switch is considered  overloaded, the orchestrator redirects the active data delivery paths to other switches.
                In \cite{leivadeas2016_dynamic}, Leivadeas et al. propose a heuristic algorithm for dynamic traffic steering across single paths in SDN enabled data centers.
                Our work is different as it considers the problem of load balancing using multiple paths.
                A closely related study to ours is the one conducted by Zhang et al. \cite{zhang2015_routing}.
                However, the authors only consider issues of load balancing specific to multicast, and ignore several important features of NFV.
                The key difference of our work compared to previous studies on the traffic engineering problem in NFV is that we take into account ECMP routing and the fundamental features of NFV such as SFC and virtualization capacity at both NFVI nodes and links.
                These factors make the problem under our consideration practical but more complex.
                In our previous work \cite{pham2016_load}, we propose an offline approximation algorithm for load balancing using multipath routing in NFV.
                In this paper, we study the optimization model and online algorithm for load balancing across multiple paths in NFV.
                In addition, we aim to provide the performance guarantee of the online algorithm in comparison with the optimal offline solution.

\section{Problem Formulation}
        \label{s-problem}

        Network Functions Virtualization is based on virtualization technologies such as those used in cloud computing. A NFV framework includes three components which are VNF, NFVI and NFV management system \cite{etsi2014_architectural}.
                In the NFVI, the network functions can be deployed as software that can run in virtual machines on the standard multi-core microprocessor architecture $\times 86$ to handle the data flow.
                One service demand of NFV may require a network service including many VNFs, in which the VNFs can be processed in a given sequence (e.g., NAT is required to be processed after the firewall function).
                When processing a service demand, the NFV management system determines the location required to deploy a VNF based on resources available on NFVI.
                Unlike single-path routing schemes, load balancing strategies split traffic among several paths in order to avoid congestion.
                The NFV system supporting load balancing both selects the routing paths and splits traffic among them under constraints on NFVI resources and the requirement of the service demand.
                With the incorporation of the SFC feature of NFV, every data flow of the demand is required to go through a sequence of VNFs dynamically allocated on NFVI nodes according to the demand requirement.
                We aim to solve the optimization problem of load balancing using ECMP in order to minimize the maximum link utilization of data flows in NFV.

        NFVI is modeled as a directed graph $G = (V,E)$, which consists of a set of nodes $V$ and directed links $E$.
                A node represents a commodity hardware device that can instantiate a VNF flexibly according to dynamic service demands.
                Let $C_{1,e}$ and $C_{2,v}$ denote respectively the bandwidth capacity of link $e \in E$ and the computing capacity of node $v \in V$.
                $F$ is a set of VNFs available on NFVI.
                We denote by $D$ a set of service demands.
                A service demand $d \in D$ is characterized by a source $s_d$, a destination $t_d$, demand volume $h_{d}$, and a service function chain $F_{d} \subset F$.
                Let $k_{vdi}$ be a parameter that equals to 1 if and only if node $v$ can provide the $i$th VNF of demand $d$.
                Throughout the paper, we will use $v$ and $t$ for a node, $e$ for a link, and $d$ for a demand unless stated otherwise.

        We define $\text{w} = \left( {{w_e}} : e \in E \right)$ to be a link metric vector of links on NFVI.
                According to service demands, a link metric vector, available system resources and ECMP routing, the system decides a flow allocation vector $\text{x}(\text{w}) = \left( x_{epd}: e \in E, d \in D, p \in P_d \right)$ where $P_d$ is a set of flows for demand $d$, and $x_{epd}$ is the traffic rate on link $e$ of flow $p$ of demand $d$ when the system uses the link metric vector $\text{w}$.
                We denote by $i_e$ and $j_e$ the starting node and terminating node of link $e$, respectively.

        Our objective is to find a link metric vector $\text{w}$ and a flow allocation vector $\text{x}(\text{w})$ in order to minimize the maximum link utilization of data flows and satisfy all requirements of service demands under constraints on NFVI resources.
                $\text{x}(\text{w})$ is determined according to ECMP routing for each $\text{w}$.

        We formulate the problem of load balancing using ECMP as a MILP model that allows us to obtain the optimal solution.
                For convenience, we summarize our notations in Table \ref{table_notation}.
	
	\begin{table}[tbp]
                \renewcommand{\arraystretch}{1.3}
                \caption{Summary of notations}
                \begin{tabular}{|>{\raggedright\arraybackslash}p{1.1cm}|>{\raggedright\arraybackslash}p{6.8cm}|} \hline
		        \multicolumn{1}{|c|}{\textbf{Notation}} & \multicolumn{1}{c|}{\textbf{Meaning}} \\ \hline
                        $V$ & The set of NFVI nodes\\ \hline
                        $E$ & The set of directed links on NFVI\\ \hline
                        $F$ & The set of VNFs available on NFVI\\ \hline
                        $D$ & The set of service demands\\ \hline
                        $n$ & The number of NFVI nodes\\ \hline
                        $m$ & The number of service demands\\ \hline
                        $C_{1,e}$ & Bandwidth capacity of link $e \in E$ \\ \hline
                        $C_{2,v}$ & Computing capacity of node $v \in V$ \\ \hline
                        $i_e$ & The starting node of link $e \in E$ \\ \hline
                        $j_e$ & The terminating node of link $e \in E$ \\ \hline
                        $h_{d}$ & The traffic volume of demand $d \in D$\\ \hline
                        $s_{d}$ & The source node of demand $d \in D$\\ \hline
                        $t_{d}$ & The destination node of demand $d \in D$\\ \hline
                        $F_{di}$ & The $i$th service function required by demand $d \in D$\\ \hline
                        $P_d$ & The set of traffic flows of demand $d$\\ \hline
                        $\chi_e$ & The total traffic rate of all data flows going through link $e$\\ \hline
                        $w_e$ & A non-negative integer variable that represents the metric of link $e$\\ \hline
                        $l_{vt}$ & A non-negative integer variable that is the length of the shortest path from node $v$ to node $t$\\ \hline
                        $x_{epd}$ & A non-negative continuous variable that represents the traffic on link $e$ of flow $p$ of demand $d$ \\ \hline
                        $g_{vt}$ & A non-negative continuous variable whose value is traffic assigned to outgoing links of node $v$ that belongs to the shortest paths from node $v$ to node $t$\\ \hline
                        $u_{et}$ & A binary variable that equals to 1 if and only if link $e$ is on a shortest path to node $t$\\ \hline
                        $b_{epd}$ & A binary variable that equals to 1 if and only if flow $p$ of demand $d$ uses link $e$\\ \hline
                        $k_{vdi}$ & A parameter that equals to 1 if and only if node $v$ can provide the $i$th VNF of demand $d$\\ \hline
                        $r$ & A non-negative continuous variable that is the maximum utilization over all links\\ \hline
                        $r_{vf}$ & The computing resources required to process function $f$ with one unit of traffic rate at node $v$\\ \hline
                        $\text{w}$  & A link metric vector of links on NFVI, $\text{w} = \left( {{w_e}} : e \in E \right)$ \\ \hline
                        $\text{x}$ & A traffic allocation vector for all demands, $\text{x} = \left( {x_{epd}: e \in E, p \in P_d, d \in D} \right)$\\ \hline
		\end{tabular}
		\label{table_notation}
	\end{table}

        The variables of our model are as follows:
        \begin{itemize}
                \item
                        ${w_e}$ is a non-negative integer variable that represents the metric of link $e$.
                \item
                        ${l_{vt}}$ is a non-negative integer variable that is the length of the shortest-path from $v$ to $t$ ($v \ne t$).
                \item
                        $x_{epd}$ is a non-negative continuous variable that represents the traffic on link $e$ of flow $p$ associated with a demand $d$.
                \item
                        $g_{vt}$ is a non-negative continuous variable whose value is traffic assigned to outgoing links of node $v$ that belongs to the shortest-paths from $v$ to $t$.
                \item
                        $u_{et}$ is a binary variable that equals to 1 if and only if link $e$ is on a shortest-path to node $t$.
                \item
                        ${b_{epd}}$ is a binary variable that equals to 1 if and only if flow $p$ of demand $d$ uses link $e$.
        \end{itemize}

        Let $\chi_e$ packet per second denote the total traffic rate on link $e$, ${\chi_e} = \sum\nolimits_{d,p} {x_{epd}}. $
	       The link utilization on each link $e$ is $\chi_e \mathord{\left/
		 {\vphantom {\chi_e C_{1,e}}} \right.
		 \kern-\nulldelimiterspace} {C_{1,e}}$.
	       The maximum utilization over all links is represented by the dependent variable
                $r = {\max _e}\left\{ {\chi_e \mathord{\left/
                 {\vphantom {{{y_e}} {{C_{1,e}}}}} \right.
                 \kern-\nulldelimiterspace} {C_{1,e}}} \right\}$.

        We now present the constraints. The conditions of flow balance at one node are given by
                \begin{alignat}{2}
                        \sum\nolimits_{\left\{ {p,e:i_e = v} \right\}} {{x_{epd}}}  - \sum\nolimits_{\left\{ {p,e:j_e = v} \right\}} {{x_{epd}}}  = 0, \nonumber\\
                                \begin{aligned}
                                        \quad {\forall d , \forall v, v \ne s_d, v \ne t_d} \label{flow-bl1}
                                \end{aligned}\\
                        \sum\nolimits_{\left\{ {p,e:i_e = s_d} \right\}} {{x_{epd}}}  = h_d,
                                \begin{aligned}
                                        \quad \forall d \label{flow-bl2}
                                \end{aligned}\\
                        \sum\nolimits_{\left\{ {p,e:j_e = t_d} \right\}} {{x_{epd}}}  = h_d,
                                        \quad \forall d \label{flow-bl3}
                \end{alignat}

        The capacity constraint on a link is
                \begin{alignat}{2}
                        \sum\nolimits_{p,d} {x_{epd}}  \leqslant {r {C_{1,e}}},
                                \begin{aligned}
                                        \quad \forall e.  \label{link-cap}
                                \end{aligned}
                \end{alignat}

        The constraint on traffic splitting according to ECMP is given by
                \begin{alignat}{2}
                        0 \leqslant {g_{{i_e} {t}}} - \sum\nolimits_{ \left\{p, d: t_d = t \right\}} {{x_{epd}}}  \leqslant \left( {1 - {u_{et}}} \right)\sum\nolimits_{ \left\{d: t_d = t \right\}} {{h_{d}}}, \nonumber \\
                                \begin{aligned}
                                        \quad \forall t, \forall e.
                                        \label{load-bl}
                                \end{aligned}
                \end{alignat}

        Constraint \eqref{load-bl} assures that if link $e$ belongs to one of the shortest-paths from node $i_e$ to node $t$ then its flow to node $t$ is equal to ${g_{i_e t}}$, a value common to all links outgoing from node $i_e$ and belonging to the shortest-paths to destination $t$.

        We express the condition of the shortest path routing as follows:
                \begin{alignat}{2}
                        \sum\nolimits_p {{x_{epd}}}  \leqslant {u_{e t_d}}{h_{d}},
                                \begin{aligned}
                                        \quad \forall d, \forall e
                                        \label{shortest1}
                                \end{aligned}
                \end{alignat}
                \begin{alignat}{2}
                        1 - {u_{e t_d}} \leqslant {l_{j_e t_d}} + {w_e} - {l_{i_e t_d}} \leqslant \left( {1 - {u_{e t_d}}} \right){M_z},
                                \begin{aligned}
                                        \quad \forall d, \forall e
                                        \label{shortest2}
                                \end{aligned}
                \end{alignat}
                where ${M_z}$ is the maximum link capacity.

        Constraint \eqref{shortest1} forces the zero flow to $t_d$ $({x_{epd}} = 0)$ in the case when link $e$ is not on the shortest-path to $t_d$.
        Constraint \eqref{shortest2} assures that if ${u_{e t_d}} = 1$, then link $e$ is on the shortest-path to $t_d$; if $u_{e t_d} = 0$, then link $e$ is not on the shortest-path to $t_d$.

        The link weights need to be larger than or equal to 1. Thus,
                \begin{alignat}{2}
                        {w_e} \geqslant 1,
                                \begin{aligned}
                                        \quad \forall e
                                        \label{weight}
                                \end{aligned}.
                \end{alignat}

        Constraints \eqref{vnf1}, \eqref{vnf2},  \eqref{flow-split1}, \eqref{flow-split2} and \eqref{flow-split3} guarantee that any flow of demand $d$ must go through its VNFs.
                \begin{alignat}{2}
                        \sum\nolimits_e {{x_{epd}}\left( {{k_{i_e di}} + {k_{j_e di}}} \right)}  > 0,
                                \begin{aligned}
                                        \quad \forall d, \forall i, \forall p, h_d > 0
                                        \label{vnf1}
                                \end{aligned}
                \end{alignat}
                \begin{alignat}{2}
                        \sum\nolimits_e {{x_{epd}}}  > 0,
                                \begin{aligned}
                                        \quad \forall d, \forall p, h_d > 0
                                        \label{vnf2}
                                \end{aligned}
                \end{alignat}
                \begin{alignat}{2}
                        0 \leqslant {x_{epd}} \leqslant {M_z}{b_{epd}},
                                \begin{aligned}
                                        \quad {\forall d, \forall e, \forall p} \label{flow-split1}
                                \end{aligned}
                \end{alignat}
                \begin{alignat}{2}
                        {x_{epd}} \geqslant \sum\nolimits_{\left\{ {e':j_{e'} = i_e} \right\}} {{x_{e'pd}}}  - {M_z}\left( {1 - {b_{epd}}} \right),
                                \begin{aligned}
                                        \quad {\forall d, \forall e, \forall p}  \label{flow-split2}
                                \end{aligned}
                \end{alignat}
                \begin{alignat}{2}
                        {x_{epd}} \leqslant \sum\nolimits_{\left\{ {e':j_{e'} = i_e } \right\}} {{x_{e'pd}}},
                                        \quad {\forall d, \forall e, \forall p}   \label{flow-split3}
                \end{alignat}

        The total computing resource required to provide VNFs for all flows through node $v$ is limited by computing resource of node $v$. We represent this constraint on node capacity  as follows:
                \begin{alignat}{2}
                        \sum\nolimits_{d,i} {{R_v}\pparen{{k_{vdi}}\sum\nolimits_{\left\{ {p,e:j_e = v} \right\}} {{x_{epd}}} ,{F_{di}}} }  \leqslant {C_{2,v}}, \quad \forall v
                                        \label{node-cap}
                \end{alignat}
        where ${R_v}\pparen{x,f} = x{r_{vf}}$, $\forall v \in V$, $\forall f \in F$.
        ${r_{vf}}$ is the amount of computing resources required to process function $f$ with one unit of traffic rate at node $v$.

        As mentioned before, we aim at minimizing the maximum utilization over all links.
        The objective function is $U\pparen{\text{w}}  = r$.
        Our MILP formulation of the load balancing problem, which allows us to obtain exact solutions, can be effectively solved for moderate network size by a MILP solver such as CPLEX \cite{cplex}.
        However, it requires the entire service demands to be known.
        In the sequel we propose an online algorithm that provides a load balancing solution for each demand on the fly.

\section{Online Solution}
        \label{s-online}
        In this section, we present the algorithm ORBIT (Online algorithm foR load BalancIng in network funcTions virtualization) which is designed to effectively address the online case of the load balancing problem in NFV.

\subsection{Algorithm Description}
        The basic idea of ORBIT is to regularly adjust a part of the traffic passing through a partition of NFVI.
                We divide NFVI into partitions in which the connection between distinct partitions is limited.
                By routing traffic through a partition, we restrict the possibility of using unnecessarily bottleneck link between partitions, thus improving network utilization and avoiding congestion.
                The detail of all steps is presented in Fig. \ref{alg-online}.

        In the preparation phase, we divide NFVI into $\kappa$ partitions with each partition including a maximum of $\epsilon n/ \kappa$ nodes while minimizing the capacity of the edges between separate partitions, in which $\kappa$ and $\varepsilon$ are the algorithm parameters.
                Particularly, we find $\Lambda \pparen{\kappa ,\varepsilon } = \left\{ {{G_i}\pparen {{V_i},{E_i}}: i = 1 \ldots \kappa } \right\}$ by solving a $\left( {\kappa ,\varepsilon } \right)$ balanced partition problem.
                We evaluate the link metric vector used as an input of ORBIT by solving the MILP formulation using the set of service demands requested previously.

        The main point of the ORBIT algorithm lies in the computation of traffic flows routed through a partition of NFVI according to the demand requirement and current situation of NFVI.
                Specifically, when a new demand arrives, we adjust traffic volume of a flow going through $G_i \in \Lambda \pparen{\kappa, \varepsilon }$ until the traffic requirement of demand $d$ is satisfied. The algorithm then distributes traffic of demand $d$ through $G_i$ according to ECMP.
                We denote by $\pi_i$ the sum of the bandwidth capacity of the edges in a minimum spanning tree of $G_i$.
                Let $q_i$ be a set of demands, in which demand $d$ is included in $q_i$, if and only if $G_i$ can provide all VNFs of $d$.
                The traffic volume of a flow going through $G_i$ depends on both the requirements of demand $d$ and the cost $\pi_i$ associated with $G_i$.
                In each adjustment, the ratio of demand bandwidth allocated to the flow through $G_i$ is computed as follows:
                \begin{alignat}{2}\label{eq-update-z}
                        {z_i} \leftarrow {z_i}\left( {1 + {1 \mathord{\left/ {\vphantom {1 {{\pi_i}\varepsilon}}} \right.\kern-\nulldelimiterspace} {{\pi_i}\varepsilon}}} \right)
                        + {1 \mathord{\left/ {\vphantom {1 {\left( { \pi_i \left| {Q(d)} \right|} \right)}}} \right. \kern-\nulldelimiterspace} {\left( {\pi_i \left| {Q(d)} \right| } \right)}}
                \end{alignat}
                where $Q\pparen{d} = \left\{ {{q_i}:d \in {q_i}} \right\}$.

        In summary, the ORBIT algorithm first divides NFVI into $\kappa$ partitions by solving a $\left( {\kappa, \varepsilon} \right)$ balanced partition problem. It then computes the traffic volume of a demand routed through a partition according to formula \eqref{eq-update-z}. A traffic flow passing through a partition is routed according to ECMP.
        We will discuss the selection of the algorithm parameters in section \ref{s-evaluation}.
        In the following section, we analyze the performance of ORBIT for online load balancing of service demands across multiple paths.

        \begin{figure}
                \begin{mdframed}
                        \begin{algorithmic}[1]
                                \MyStatex{\textbf{Data:} NFVI $G = \pparen{V,E}$, $C_{1,e}$, $C_{2,v}$, $\text{w}$, $\kappa$, $\varepsilon$ }
                                \MyStatex{\textbf{Result:} Multipath routing solution upon demand arrival}
                                \MyState{find $\Lambda \pparen{\kappa ,\varepsilon } = \left\{ {{G_i}\pparen {{V_i},{E_i}}: i = 1 \ldots \kappa } \right\}$ by solving a $\left( {\kappa ,\varepsilon } \right)$ balanced partition problem}
                                \MyState{initialize $Q = \left\{ {{q_i}:{q_i} = \emptyset, i = 1 \ldots \kappa } \right\}$}
                                \MyState{initialize $z_i = 0$, $i = 1 \ldots \kappa$}
                                \While{true}
                                        \If{$<$a new demand $d$ arrives$>$}
                                                \ForAll{$G_i \in \Lambda \pparen{\kappa ,\varepsilon } $}
                                                        \If{$<$$G_i$ can provide all VNFs of $d$$>$}
                                                                \MyState{${q_i} = {q_i} \cup \left\{ d \right\}$}
                                                        \EndIf
                                                \EndFor
                                                \MyState{$Q\pparen{d} = \left\{ {{q_i}:d \in {q_i}} \right\}$}
                                                \If{$Q\pparen{d} =  \emptyset$}
                                                        \MyState{Reject demand $d$ and go to line $4$}
                                                \EndIf
                                                \While {$\sum\nolimits_{i : q_i \in Q(d)} {{z_i}}  < 1$}
                                                        \ForAll{$q_i \in Q(d)$}
                                                                \MyState{${z_i} \leftarrow {z_i}\left( {1 + {1 \mathord{\left/ {\vphantom {1 {{\pi_i \varepsilon}}}} \right.\kern-\nulldelimiterspace} {{\pi_i \varepsilon}}}} \right)
                                                                + {1 \mathord{\left/ {\vphantom {1 {\left( {{\pi_i} \left| {Q(d)} \right|} \right)}}} \right. \kern-\nulldelimiterspace} {\left( {{\pi_i} \left| {Q(d)} \right|} \right)}}$}
                                                        \EndFor
                                                \EndWhile
                                                \If{$<$link capacity is satisfied$>$}
                                                        \MyState{distribute traffic ${{{h_d z_i}} \mathord{\left/
         {\vphantom {{{h_d z_i}} {\sum\nolimits_{i: q_i \in Q(d)} {{z_i}} }}} \right.
         \kern-\nulldelimiterspace} {\sum\nolimits_{i: q_i \in Q(d)} {{z_i}} }}$ through $G_i$ according to ECMP}
                                                        \MyState{update capacity of $G_i$ and link capacity}
                                                \Else
                                                        \MyState{Reject demand $d$}
                                                \EndIf
                                        \EndIf
                                \EndWhile
                        \end{algorithmic}
                \end{mdframed}
                \caption{The ORBIT algorithm}
                \label{alg-online}
        \end{figure}

\subsection{Theoretical Analysis}

        In order to analyze the algorithm performance, we consider the following minimization problem as the primal program $\mathcal{P}$, which is given by
                \begin{alignat}{2}
                        \text{Minimize } \quad
                                &\sum\nolimits_{i = 1 \ldots \kappa } {{\pi _i}{z_i}} \label{primal_obj} \\
                        \text{Subject to: } \quad
                                &\sum\nolimits_{i \in Q(d)} {{z_i}} \geqslant 1, \quad \forall d \in D                                 \label{primal_constraint1}\\
                                &{z_i}  \geqslant 0, \quad \forall 1 \leqslant i \leqslant \kappa
                \end{alignat}

        The dual problem $\mathcal{D}$ of the primal problem $\mathcal{P}$  is as follows:
                \begin{alignat}{2}
                        \text{Maximize } \quad
                                &\sum\nolimits_{d \in D} {{\zeta _d}} \nonumber \\
                        \text{Subject to: } \quad
                                &\sum\nolimits_{d:i \in Q(d)} {{\zeta_d}}  \leqslant {\pi_i},
                                        \quad \forall 1 \leqslant i \leqslant \kappa \label{dual_constraint1}\\
                                &\quad {\zeta_d}  \geqslant 0,
                                        \quad \forall d \in D
                \end{alignat}

        In the online version of $\mathcal{P}$ (i.e., the online primal program), the constraints are given to the algorithm one-by-one. In the online version of $\mathcal{D}$ (i.e., the online dual program), a new variable $\zeta_d$ is introduced and the set of constraints in which $\zeta_d$ appears is also updated when a new demand $d$ arrives.
        The performance of an online solution is represented by the competitive ratio that is the ratio between the cost of the online solution and that of the optimal offline solution.
        Lemma 1 to 3 will give the foundation to derive the competitive ratio of ORBIT.

	\begin{lemma}
	       \label{lemma-feasible}
                The ORBIT algorithm produces a feasible solution for the online primal load balancing problem.
	\end{lemma}
        \begin{IEEEproof}
                Consider a constraint $\sum\nolimits_{i \in Q(d)} {{z_i}}  \geqslant 1$ in $\mathcal{P}$. When demand $d$ arrives, ORBIT increases the values of the variables $z_i$ until the constraint is satisfied. Hence, the solution produced by ORBIT is feasible, which demonstrates the claim.
        \end{IEEEproof}

        Lemma \ref{lemma-constraint} presents a feasible solution provided by ORBIT for the online version of $\mathcal{D}$. We add ${\zeta_d} \leftarrow {\zeta_d} + 1$ between line 18 and 19 in the ORBIT algorithm in order to find a solution for $\mathcal{D}$.

	\begin{lemma}
                \label{lemma-constraint}
                A feasible solution for the online dual load balancing problem is obtained by dividing the online dual solution which ORBIT produces by $c \varepsilon \log \kappa$ where $c$ is a constant. The cost of the feasible solution is $D_{f} = D_o/(c \varepsilon \log \kappa)$.
	\end{lemma}

        \begin{IEEEproof}
                We first prove by induction that, for $\forall d \in D$,
                \begin{equation}\label{eq_eval_dual_constraint}
                        {z_i} \geqslant \frac{1}{\kappa }\left( {{{\left( {1 + \frac{1}{{{\pi _i}\varepsilon }}} \right)}^{\sum\nolimits_{d:i \in Q(d)} {{\zeta _d}} }} - 1} \right).
                \end{equation}

                Initially, $z_i = 0$ and $\zeta_d = 0$, so \eqref{eq_eval_dual_constraint} is true.
                Let consider an iteration in which $\zeta_k$ increases by 1. We denote by $z_{i,1}$ and  $z_{i,2}$ the values of $z_{i}$ before and after the increment, respectively. Using \eqref{eq-update-z}, we have
                \begin{align}
                        {z_{i,2}} &= {z_{i,1}}\left( {1 + \frac{1}{{{\pi _i}\varepsilon }}} \right) + \frac{1}{{{\pi _i}\left| {Q(d)} \right|}} \nonumber\\
                        &= \frac{1}{\kappa }\left( {{{\left( {1 + \frac{1}{{{\pi _i}\varepsilon }}} \right)}^{\sum\nolimits_{d:i \in Q(d)\backslash \left\{ k \right\}} {{\zeta _d}} }} - 1} \right) \nonumber\\
                        &\quad \times \left( {1 + \frac{1}{{{\pi _i}\varepsilon }}} \right) + \frac{1}{{{\pi _i}\left| {Q(d)} \right|}}.
                         \label{eq_eval_dual_constraint2}
                \end{align}

                Since $\left| {Q(d)} \right| \leqslant \kappa$, we find
                 \begin{equation}
                        {z_{i,2}} \geqslant \frac{1}{\kappa }\left( {{{\left( {1 + \frac{1}{{{\pi _i}\varepsilon }}} \right)}^{\sum\nolimits_{d:i \in Q(d)} {{\zeta _d}} }} - 1} \right).
                 \end{equation}
                Thus, \eqref{eq_eval_dual_constraint} is true for $\forall d \in D$.

                Due to the fact that the algorithm never increases $z_i$ if $z_i \geqslant 1$, combining $\left| {{Q(d)}} \right| \geqslant 1$, $\pi_i \geqslant 1$ and \eqref{eq-update-z}, we have $z_i \leqslant 3$. From this inequality and \eqref{eq_eval_dual_constraint}, we find
                 \begin{equation}\label{eq_eval_dual_constraint3}
                        3 \geqslant \frac{1}{\kappa }\left( {{{\left( {1 + \frac{1}{{{\pi _i}\varepsilon }}} \right)}^{\sum\nolimits_{d:i \in Q(d)} {{\zeta _d}} }} - 1} \right).
                 \end{equation}

                Using the fact that $\varepsilon \geqslant 1$, $\pi_i \geqslant 1$, simplifying \eqref{eq_eval_dual_constraint3}, we obtain
                \begin{equation}\label{eq_eval_dual_constraint4}
                        \sum\nolimits_{d:i \in Q(d)} {{\zeta _d}}  \leqslant \log \left( {3\kappa  + 1} \right)\left( {1 + {\pi _i}\varepsilon } \right) = {\pi _i}O\left( {\varepsilon \log \kappa } \right).
                 \end{equation}

                From \eqref{eq_eval_dual_constraint4}, dividing the solution by $c \varepsilon \log \kappa$ where $c$ is a constant, we obtain a feasible solution for the online dual load balancing problem with the cost $D_{f} = D_o/(c \varepsilon \log \kappa)$, which proves the claim.
        \end{IEEEproof}

        Let $P_o$ and $D_o$ be the values of the objective function of the solutions that ORBIT produces for the online primal and dual programs, respectively.
                Let $\Delta P_o$ and $\Delta D_o$ be the changes of $P_o$ and $D_o$ in an iteration of ORBIT.
                Lemma \ref{lemma-profit} gives a comparison between $\Delta P_o$ and $\Delta D_o$.

	\begin{lemma}
	\label{lemma-profit}
                In each iteration of the ORBIT algorithm, the relationship between the online primal and dual solutions that ORBIT produces is given by $\Delta P_o \leqslant 2 \Delta D_o$.
	\end{lemma}

        \begin{IEEEproof}
                From the cost function \eqref{primal_obj} of the primal program and the adjustment function \eqref{eq-update-z}, we have
                \begin{align}
                        \Delta P_o &= \sum\nolimits_{i:{q_i} \in Q(d)} {{\pi _i}\left( {\frac{{{z_i}}}{{{\pi _i}\varepsilon}}
                        + \frac{1}{{{\pi _i}\left| {Q(d)} \right|}}} \right)}  \nonumber\\
                        &= \sum\nolimits_{i:{q_i} \in Q(d)} {\left( {\frac{{{z_i}}}{{\varepsilon}} + \frac{1}{{\left| {Q(d)} \right|}}} \right)}
                         \label{eq_cost_change}
                \end{align}

                Since constraint \eqref{primal_constraint1} is infeasible when ORBIT updates the primal and dual solutions, we find $\Delta P_o \leqslant 2$. In addition, the change in the dual cost is 1. Thus, we obtain $\Delta P_o \leqslant 2 \Delta D_o$, which proves the claim.
        \end{IEEEproof}

        We are almost ready to derive the competitive ratio of the ORBIT algorithm that is introduced in the following proposition.

	\begin{proposition}
	\label{thm-o-perf}
                ORBIT produces an online load balancing solution which is $O\pparen{\varepsilon \log \kappa}$-competitive.
	\end{proposition}

        \begin{IEEEproof}
                First, by Lemma \ref{lemma-feasible} the online solution produced by the ORBIT algorithm is feasible.
                From Lemma \ref{lemma-profit} and the fact that initially $P_o = D_o = 0$, we find that the ORBIT algorithm produces online primal and dual solutions such that $P_o \leqslant 2 D_o$.
                By Lemma \ref{lemma-constraint}, we get a feasible solution for the online dual program with value $D_f = D_o/(c \varepsilon \log \kappa)$ where $c$ is a constant.
                Therefore, we find $P_o \leqslant 2 D_f (c \varepsilon \log \kappa)$.
                Using this inequality and the weak duality theorem, we obtain that the primal solution is at most $2 c \varepsilon \log \kappa$ times the optimal primal solution, which demonstrates the proposition.
        \end{IEEEproof}

        \begin{figure*}[!t]
                \centering
                \subfloat[\label{fig-epsilon-i2-k2-link}Internet2, $\kappa=2$]{
                        \centering
                        \includegraphics[width=0.4\columnwidth]{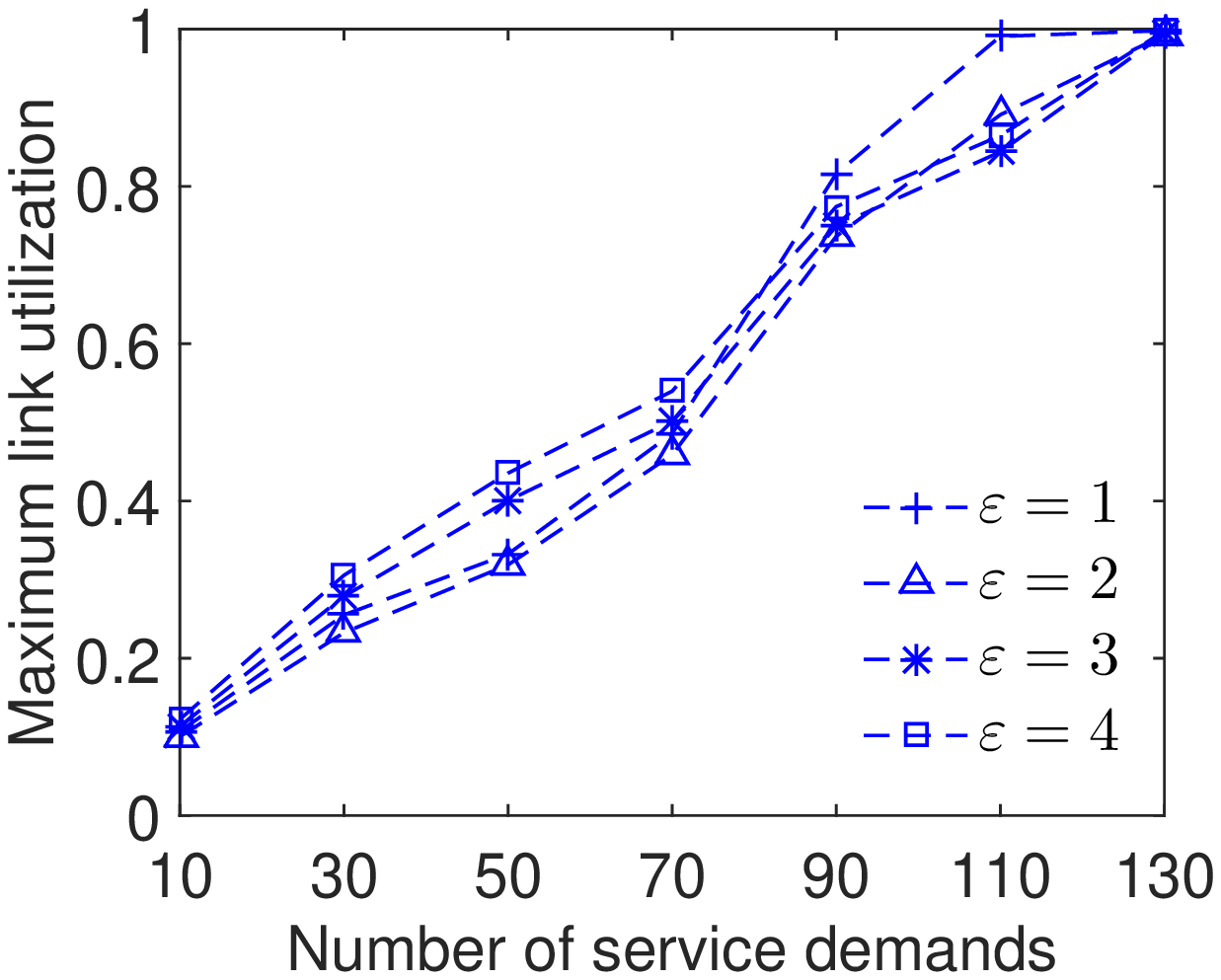}
                }
                ~
                \subfloat[\label{fig-epsilon-i2-k2-accept}Internet2, $\kappa=2$]{
                        \centering
                        \includegraphics[width=0.4\columnwidth]{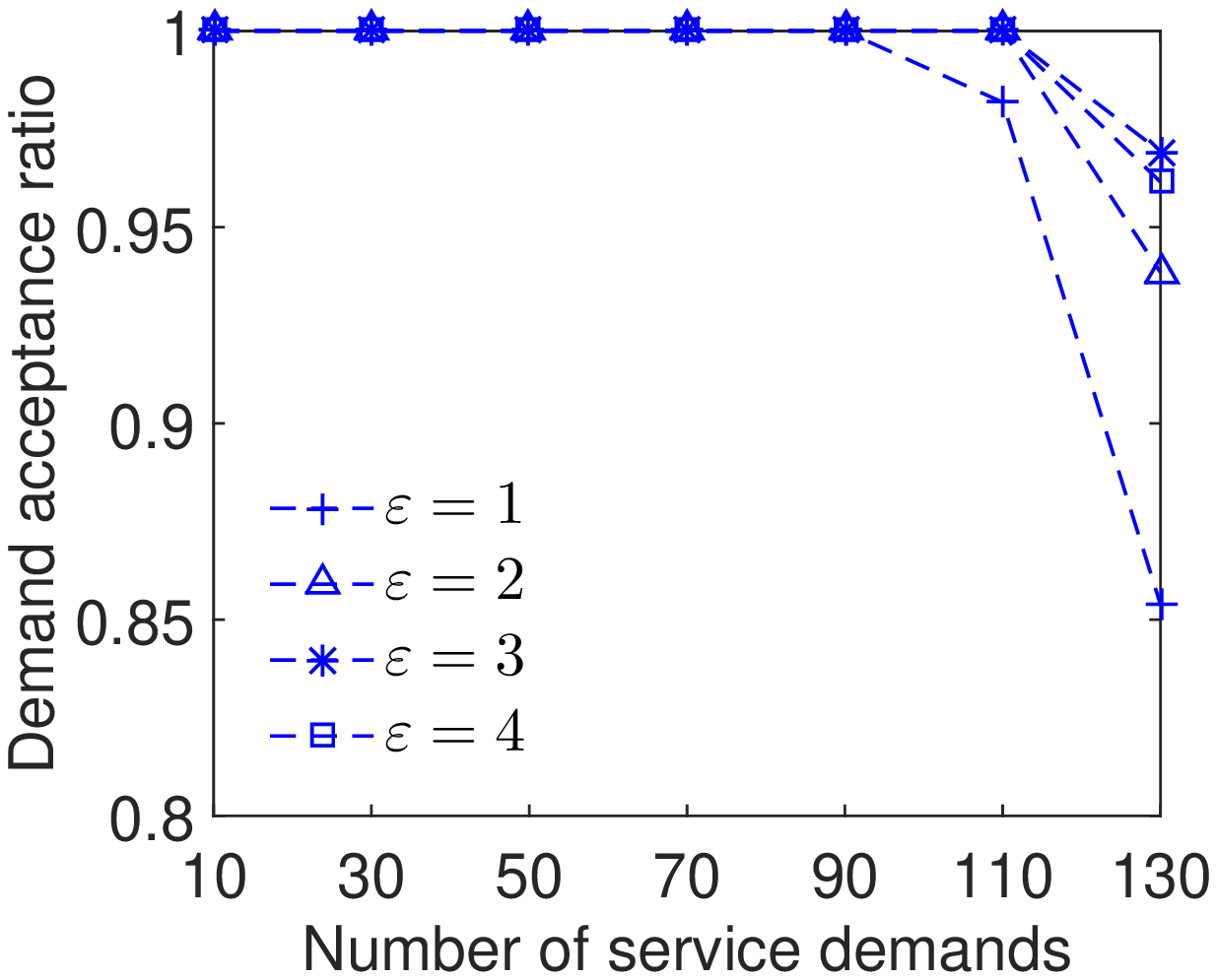}
                }
                ~
                \subfloat[\label{fig-epsilon-i2-k3-link}Internet2, $\kappa=3$]{
                        \centering
                        \includegraphics[width=0.4\columnwidth]{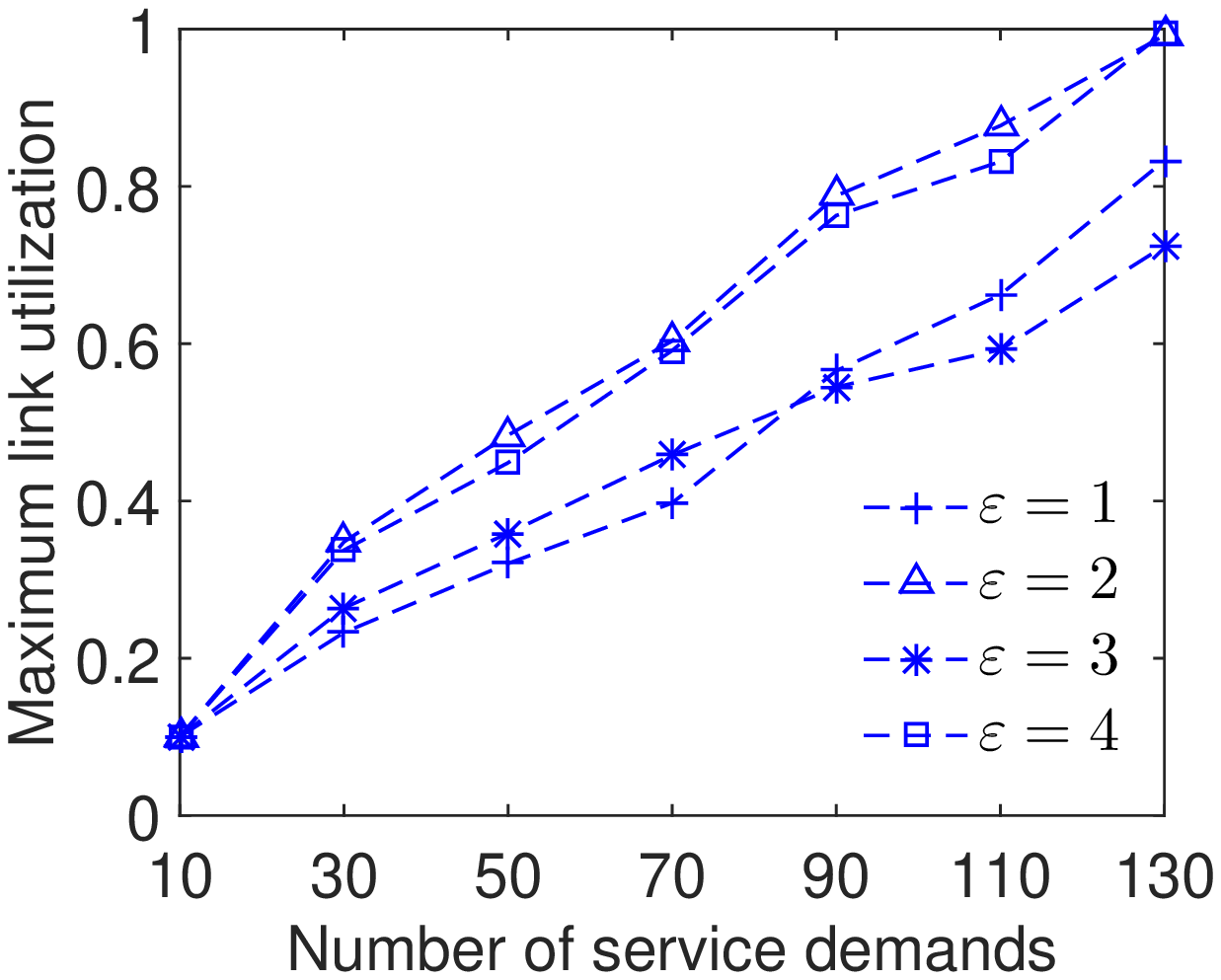}
                }
                ~
                \subfloat[\label{fig-epsilon-i2-k3-accept}Internet2, $\kappa=3$]{
                        \centering
                        \includegraphics[width=0.4\columnwidth]{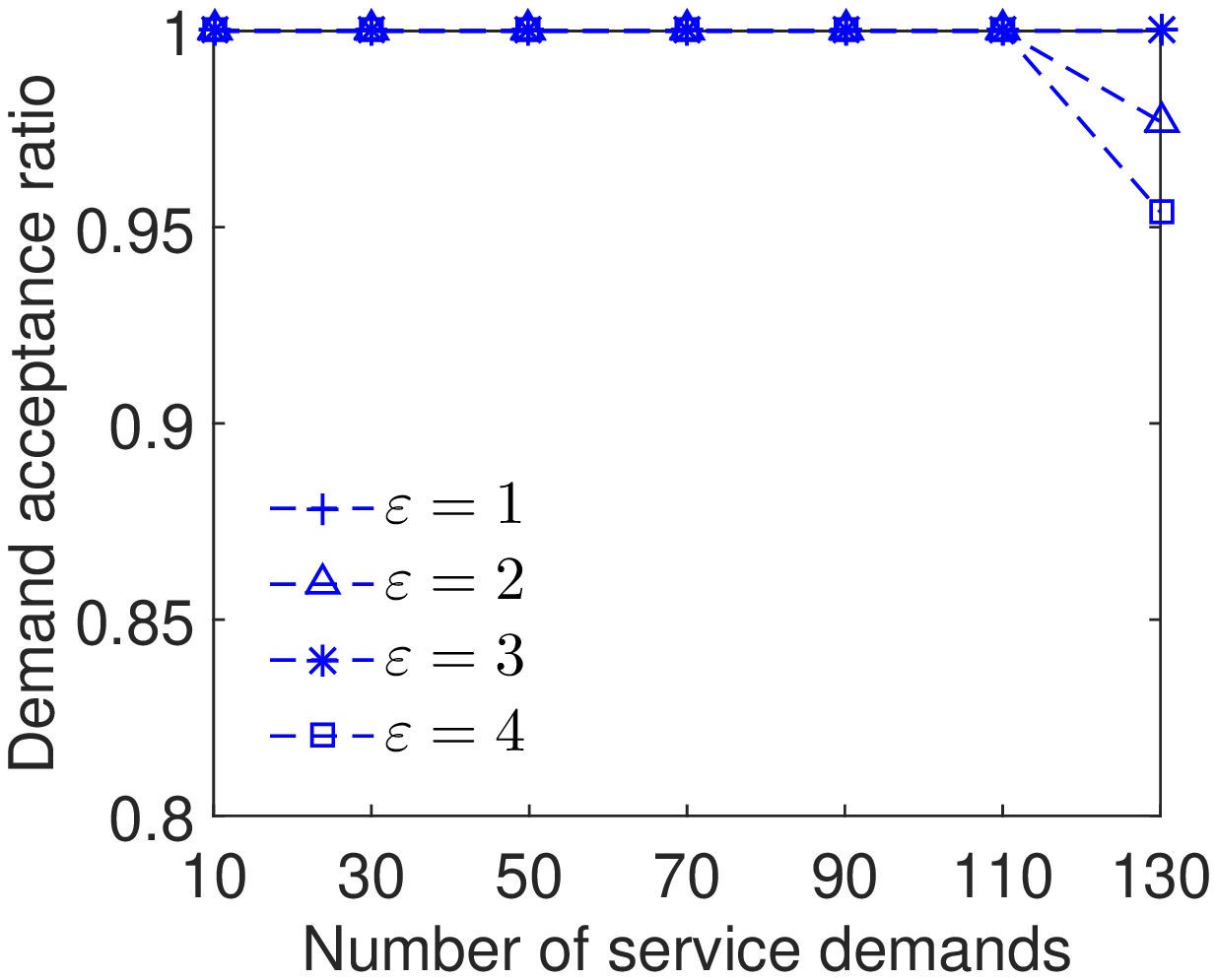}
                }
                \\
                \subfloat[\label{fig-epsilon-g-k2-link}Geant, $\kappa=2$]{
                        \centering
                        \includegraphics[width=0.4\columnwidth]{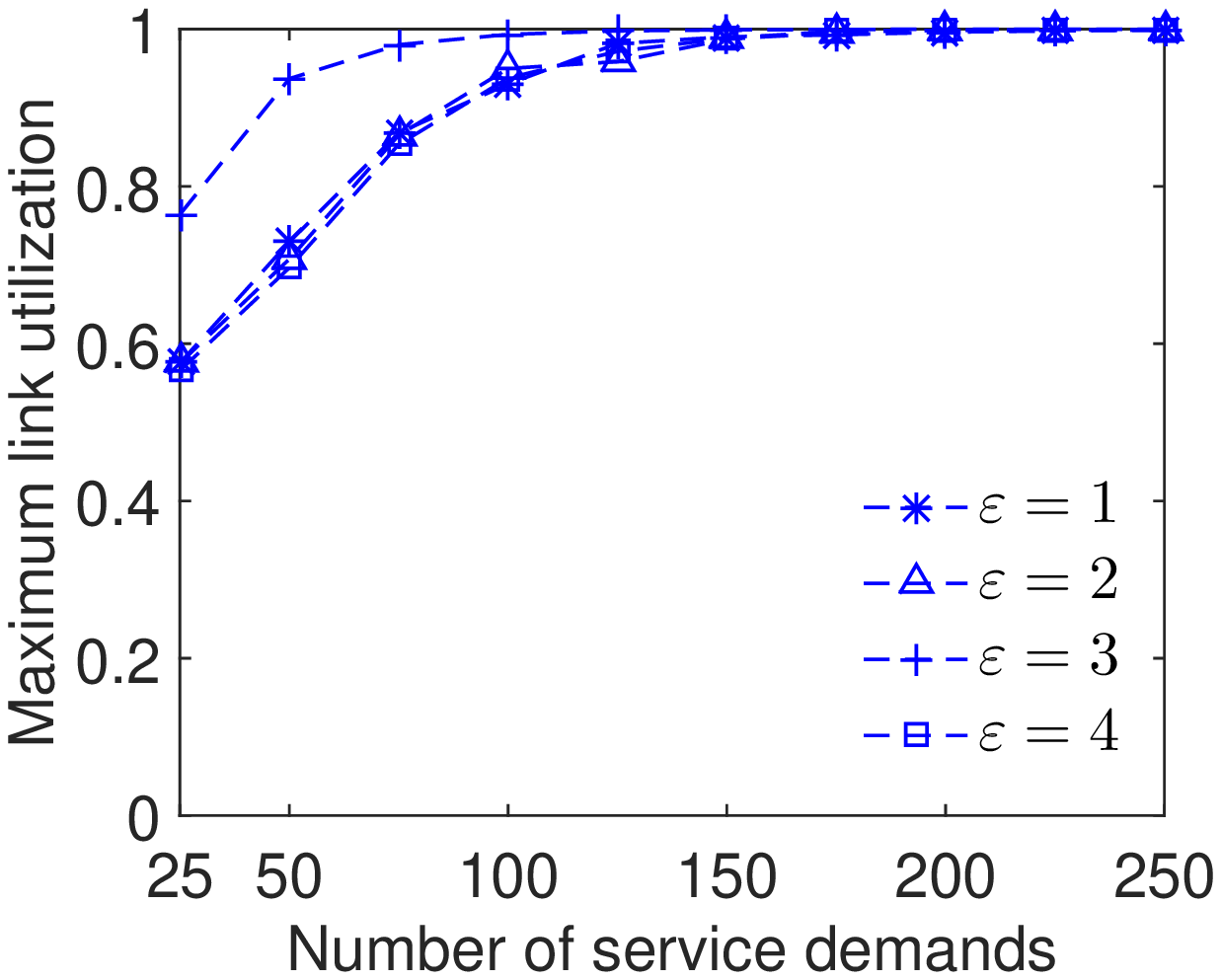}
                }
                ~
                \subfloat[\label{fig-epsilon-g-k2-accept}Geant, $\kappa=2$]{
                        \centering
                        \includegraphics[width=0.4\columnwidth]{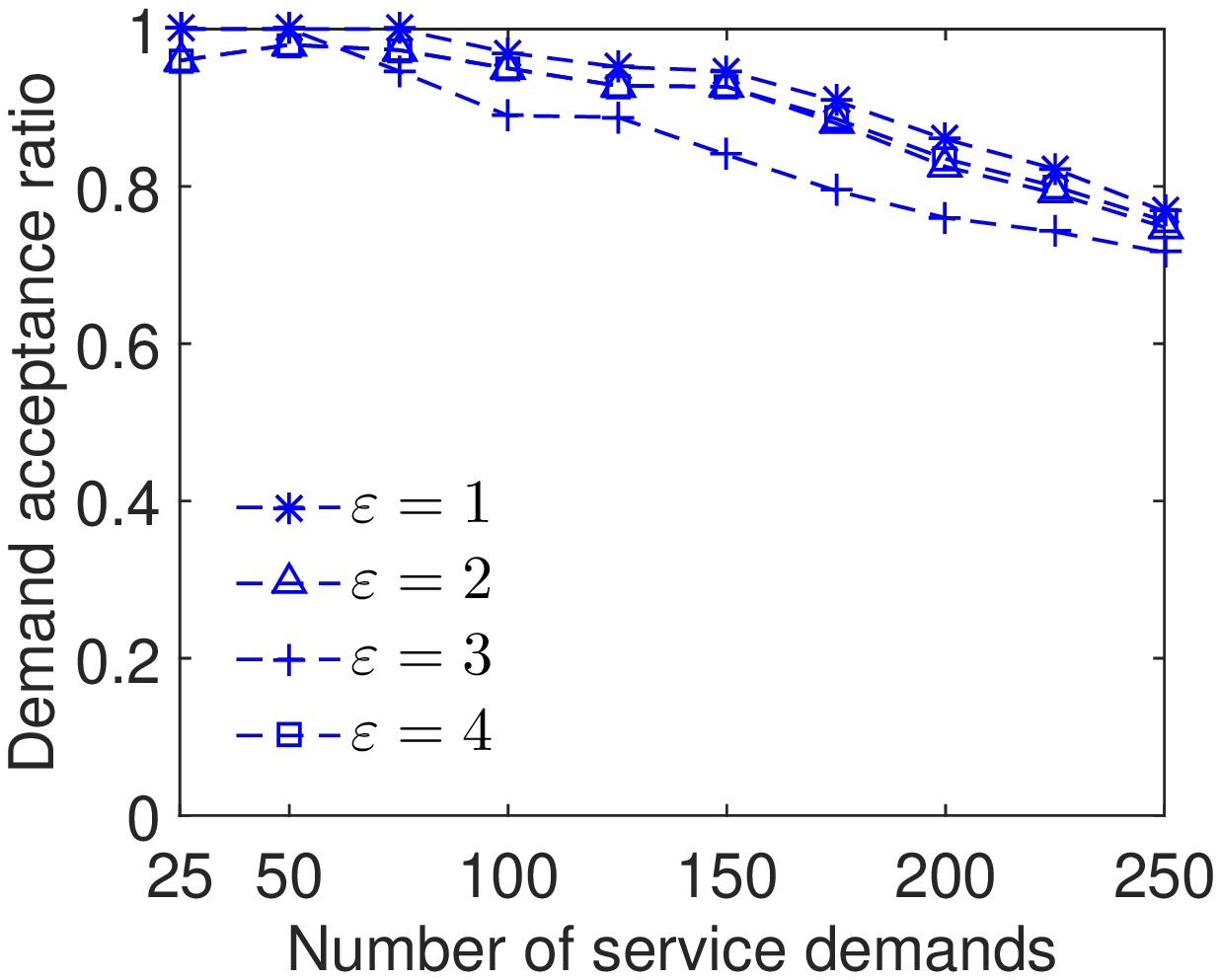}
                }
                ~
                \subfloat[\label{fig-epsilon-g-k3-link}Geant, $\kappa=3$]{
                        \centering
                        \includegraphics[width=0.4\columnwidth]{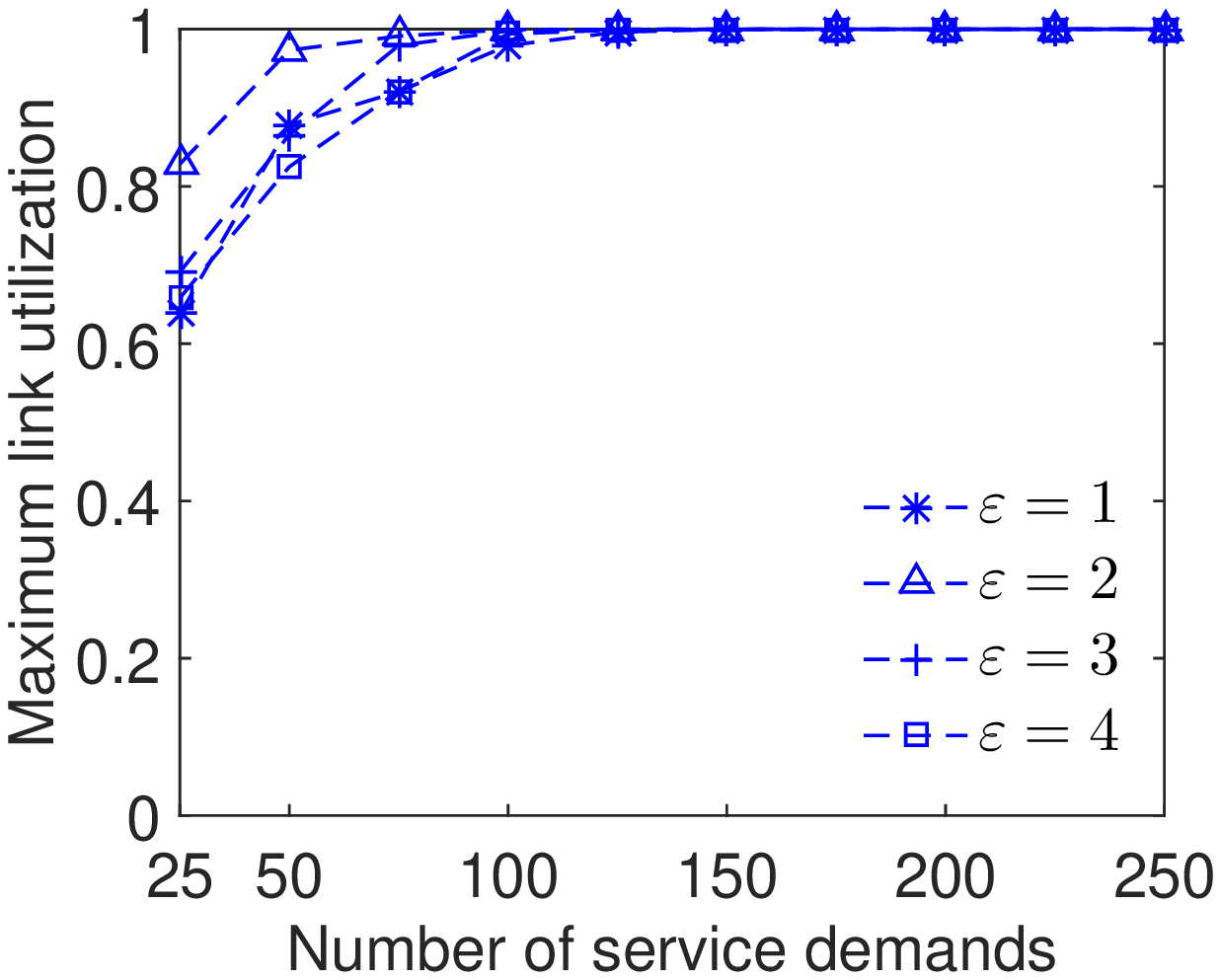}
                }
                ~
                \subfloat[\label{fig-epsilon-g-k3-accept}Geant, $\kappa=3$]{
                        \centering
                        \includegraphics[width=0.4\columnwidth]{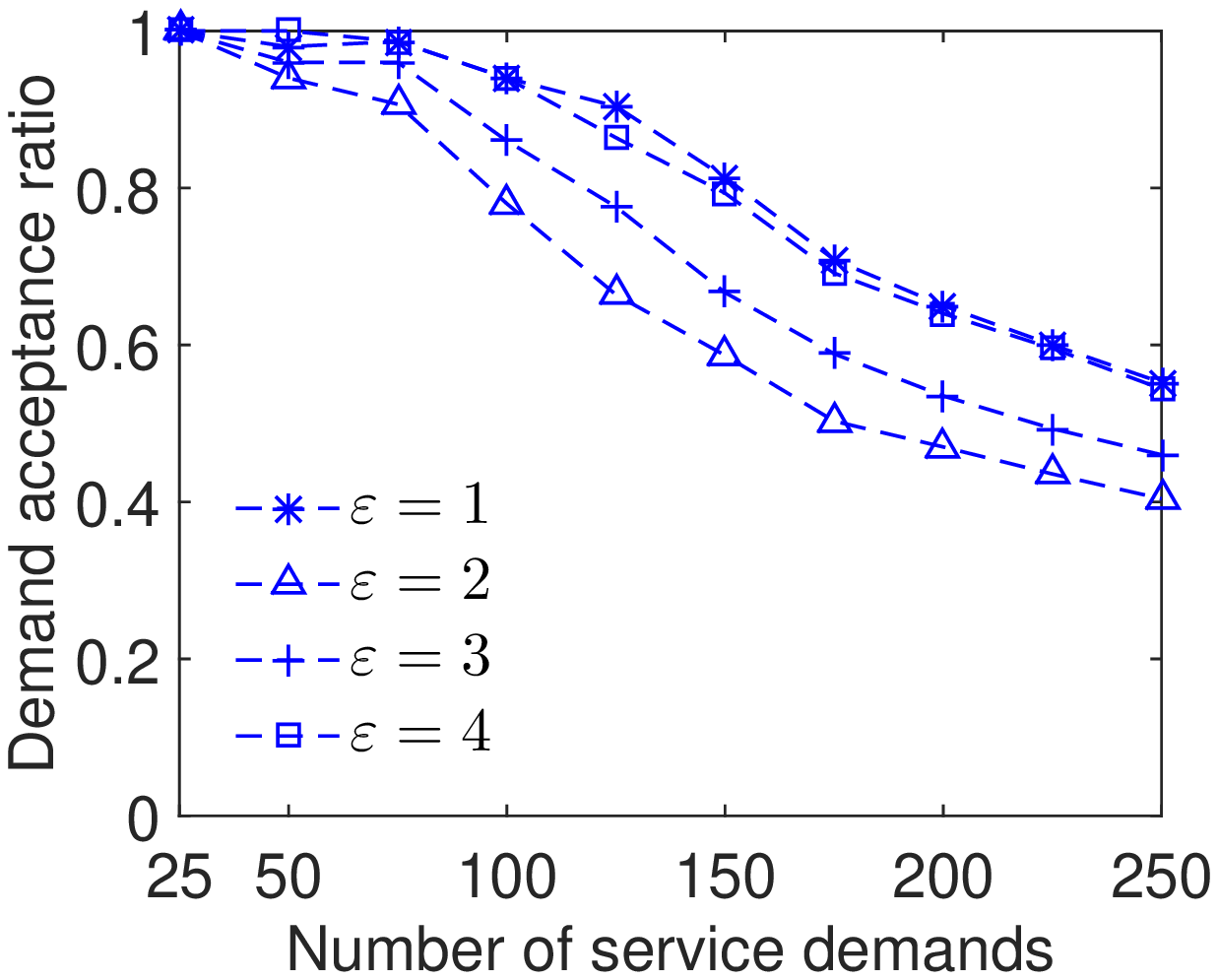}
                }
                \caption{Impact of parameter $\varepsilon$ on the algorithm performance}
                \label{fig-epsilon}
        \end{figure*}

        \begin{figure}[!t]
                \centering
                \subfloat[\label{fig-kappa-i2-link}]{
                        \centering
                        \includegraphics[width=0.4\columnwidth]{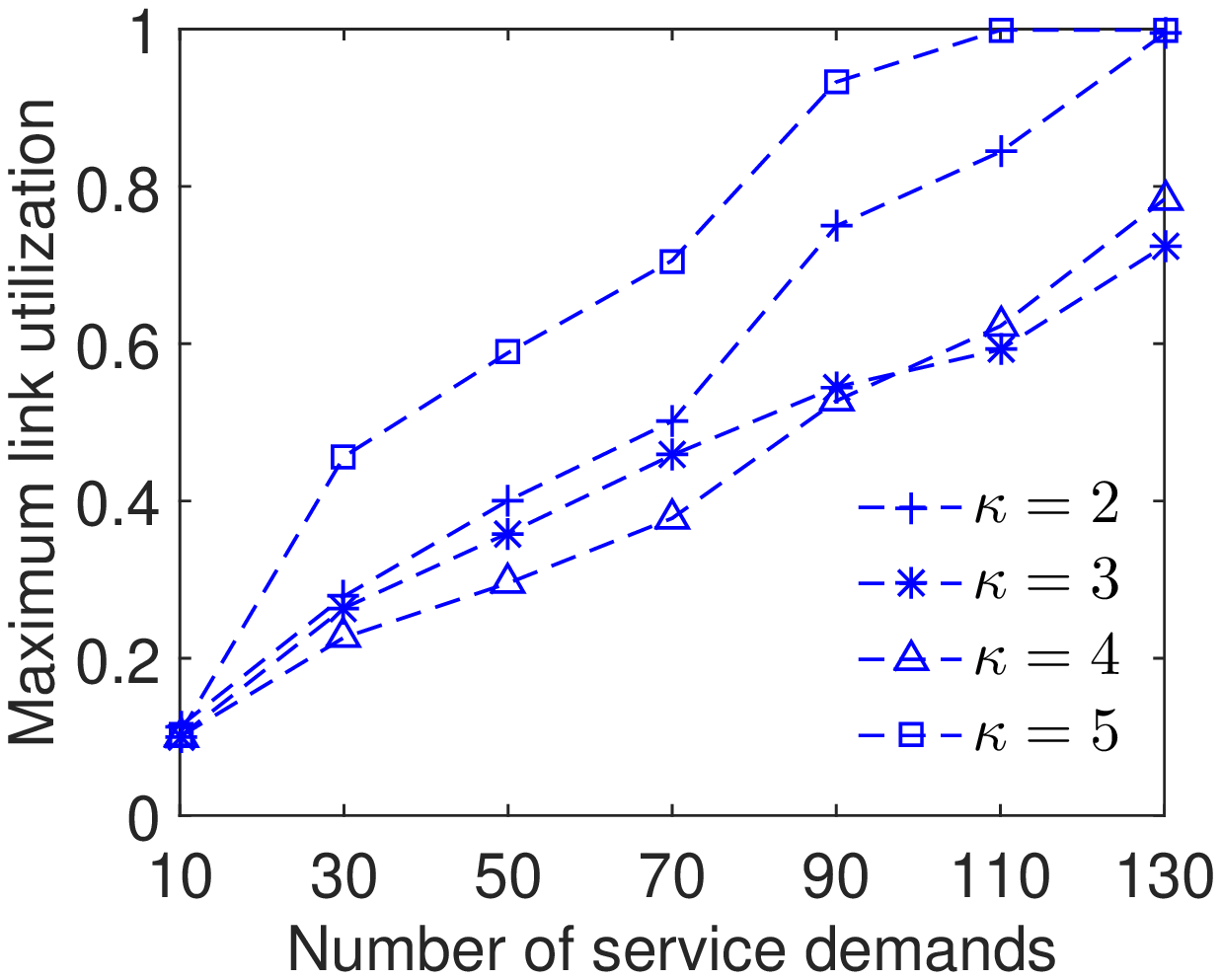}
                }
                ~
                \subfloat[\label{fig-kappa-i2-accept}]{
                        \centering
                        \includegraphics[width=0.4\columnwidth]{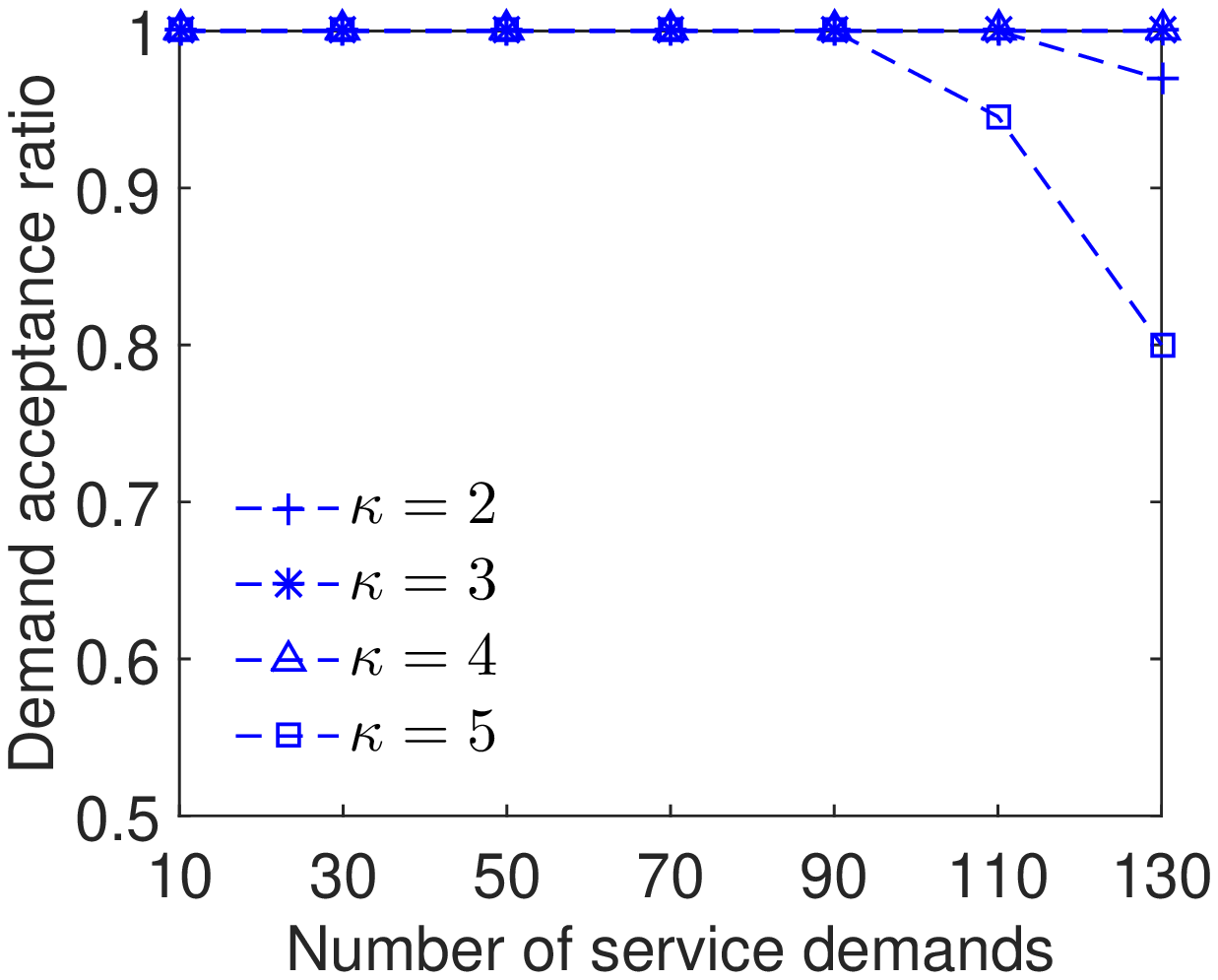}
                }
                \caption{Impact of parameter $\kappa$ on the algorithm performance using Internet2}
                \label{fig-kappa-i2}
        \end{figure}

        \begin{figure}[!t]
                \centering
                \subfloat[\label{fig-kappa-g-link}]{
                        \centering
                        \includegraphics[width=0.4\columnwidth]{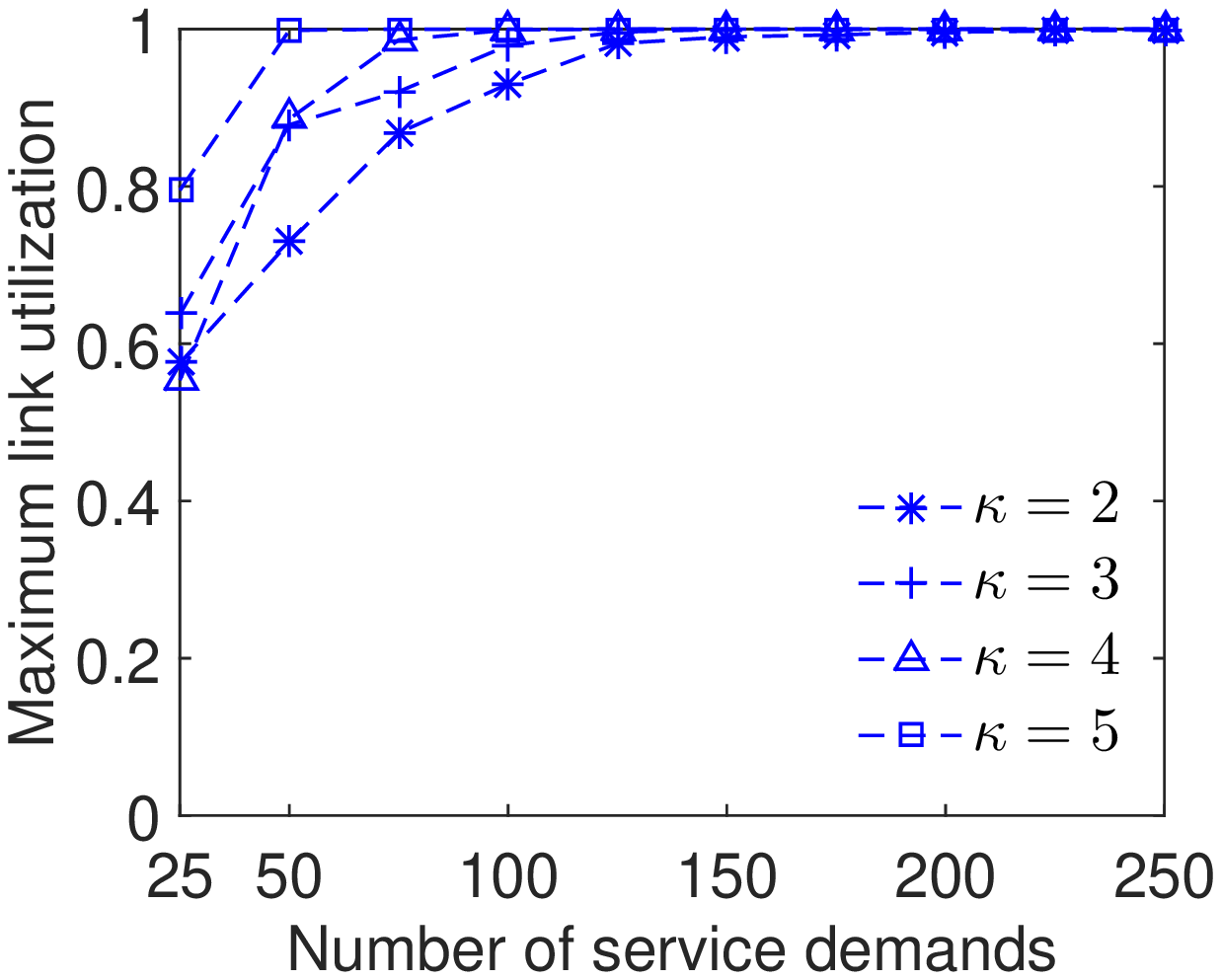}
                }
                ~
                \subfloat[\label{fig-kappa-g-accept}]{
                        \centering
                        \includegraphics[width=0.4\columnwidth]{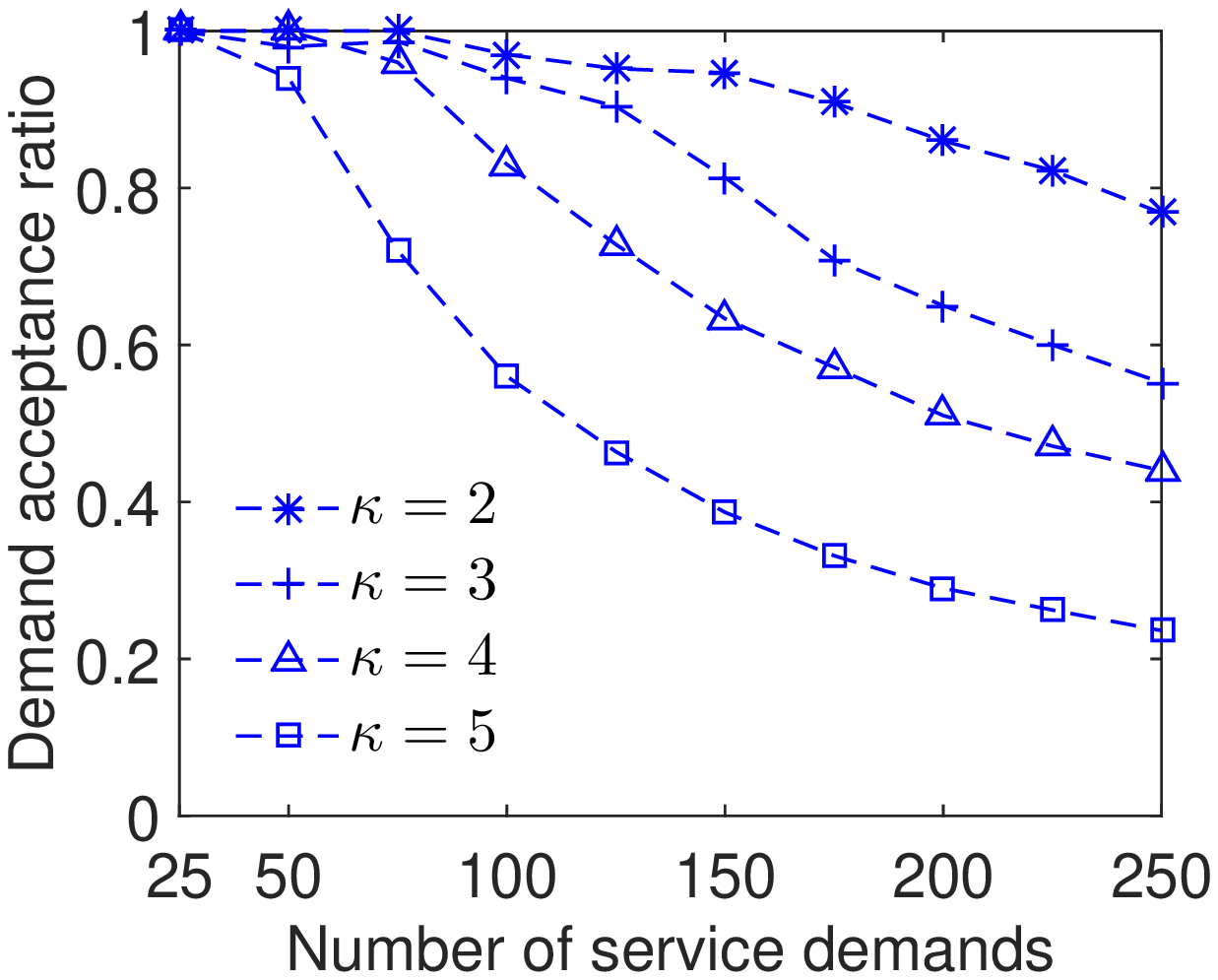}
                }
                \caption{Impact of parameter $\kappa$ on the algorithm performance using Geant}
                \label{fig-kappa-g}
        \end{figure}

\section{Evaluation}
        \label{s-evaluation}

        In this section, we describe the setting of our experiments using two real-world datasets. We then analyze the experimental results including the impact of the algorithm parameters and the evaluation of algorithm performance in comparison with the optimal and offline solutions.

        In our experiments, we use two real-world datasets that contain the network topology and traffic matrices measured at different times.
                The first dataset is the Internet2 research network including a topology of 12 nodes, 30 links and traffic matrices with 130 demands \cite{internet2}.
                The traffic traces are recorded for a duration of 30 minutes.
                The second dataset is the Geant dataset that contains a topology of 22 nodes and 72 links and traffic matrices with 250 demands.
                The traffic volume of demands in our experiments is a half of total traffic traces over four months in the Geant network.
                In our experiments, we consider four VNFs available on NFVI. The computing capacity, the resource requirements of a VNF, and the SFC of each demand are randomly generated.
                In the experiments, we solve the MILP formulation for 10 service demands to compute a weight system for ORBIT.
                In the preparation phase of ORBIT, we use a multilevel graph partitioning algorithm for dividing NFVI into $\kappa$ partitions \cite{hendrickson1995_multilevel}.

        We first evaluate the performance of ORBIT when varying the value of the algorithm parameters in order to gain insight into the parameter setting of ORBIT in practice.
                Fig. \ref{fig-epsilon} shows the maximum link utilization and the demand acceptance ratio obtained by ORBIT for various $\varepsilon$ when $\kappa=2$ and $\kappa=3$ in the experiments using the two datasets.
                We observe that the best value of $\varepsilon$ for ORBIT is not affected by the number of partitions $\kappa$.
                Specifically, ORBIT obtains better results when $\varepsilon=3$ and $\varepsilon=1$ in the experiments using dataset Internet2 and Geant respectively.
                Thus, we use these values of $\varepsilon$ for other experiments.

        Fig. \ref{fig-kappa-i2} and \ref{fig-kappa-g} show the maximum link utilization and the demand acceptance ratio obtained by ORBIT for various $\kappa$ and the two datasets.
                We observe that ORBIT achieves better results when $\kappa$ increases until a threshold is reached, beyond which further increase in $\kappa$ degrades performance.
                Particularly, the algorithm obtains the best performance in both the maximum link utilization and the demand acceptance ratio when $\kappa = 3$ in the experiment using dataset Internet2 and $\kappa = 2$ in the experiment using dataset Geant.
                The reason is that ORBIT balances traffic over the partitions according to not only the number of partitions, but also the resource capacity of a partition.
                Under a high value of $\kappa$, ORBIT is likely prevented from distributing a flow through a small partition due to the lack of resource, resulting in worse performance.
                When we choose a different set of traffic demands in both datasets, the performance improvement also occurs at values of $\kappa$ between 2 and 3.
                It suggests that we can obtain an appropriate value of $\kappa$ for a specific NFVI by observing the algorithm behaviour in a time duration.

\begin{figure}[!t]
                \centering
                \subfloat[\label{fig-compare-opt-b}Internet2]{
                        \centering
                        \includegraphics[width=0.4\columnwidth]{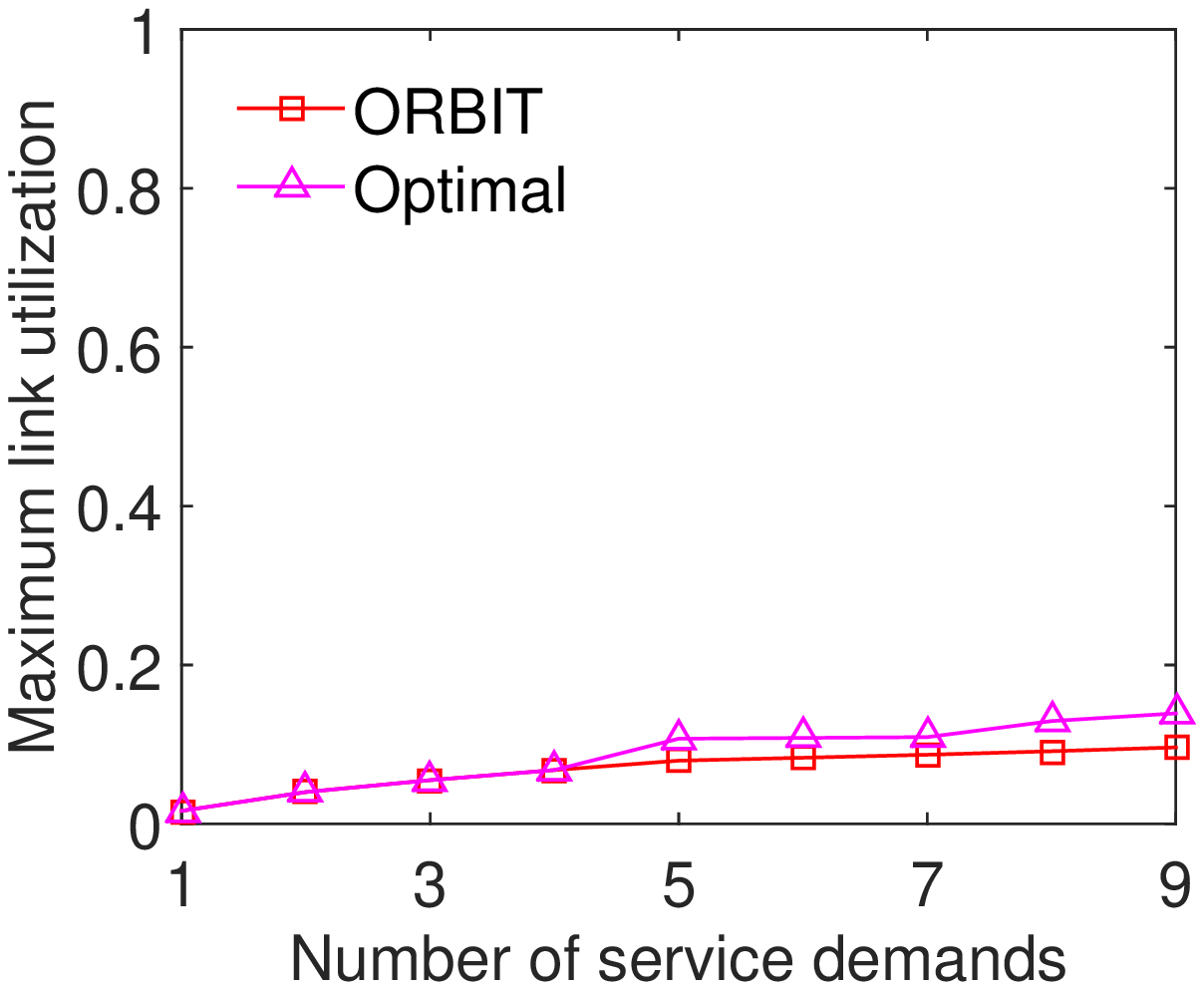}
                }
                ~
                \subfloat[\label{fig-compare-opt-c}Geant]{
                        \centering
                        \includegraphics[width=0.4\columnwidth]{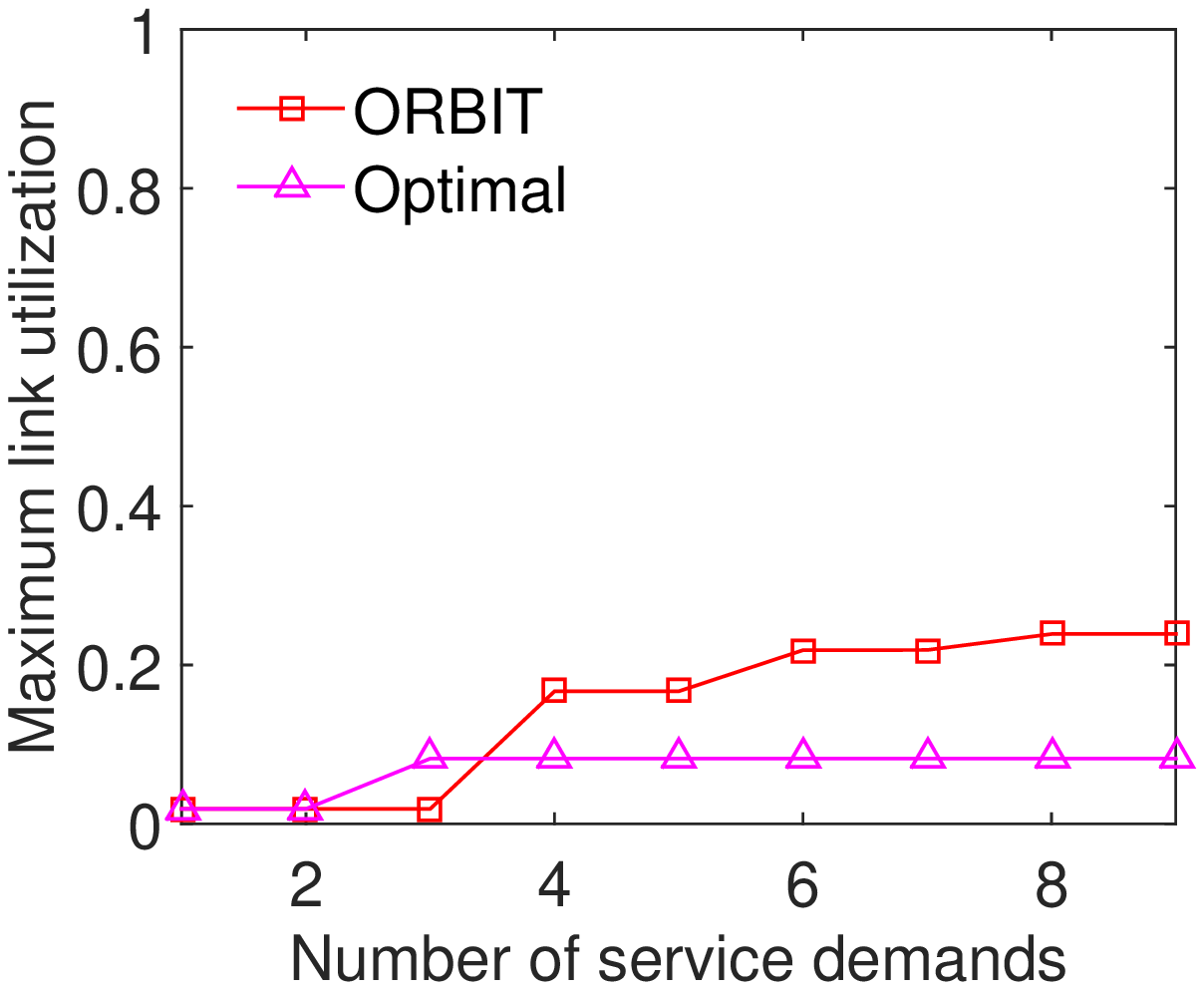}
                }
                \caption{Comparison between ORBIT and the optimal solution}
                \label{fig-compare-opt}
        \end{figure}

        Second, we analyze the performance of the ORBIT algorithm in comparison with the optimal solution.
                We use the IBM ILOG CPLEX Optimizer to solve the MILP model in order to obtain the optimal results.
                Fig. \ref{fig-compare-opt} provides a comparison of the maximum link utilization obtained by ORBIT and the optimal solution.
                In both datasets, the results provided by ORBIT are close to the optimal solution.
                Specially, we observe that ORBIT obtains slightly better results in the experiments using the Geant dataset.
                This occurs because when NFVI is divided into several partitions, ORBIT can use both the shortest paths and many near-optimal paths.

\begin{figure}[!t]
                \centering
                \subfloat[\label{fig-compare-offline-i2-link}]{
                        \centering
                        \includegraphics[width=0.4\columnwidth]{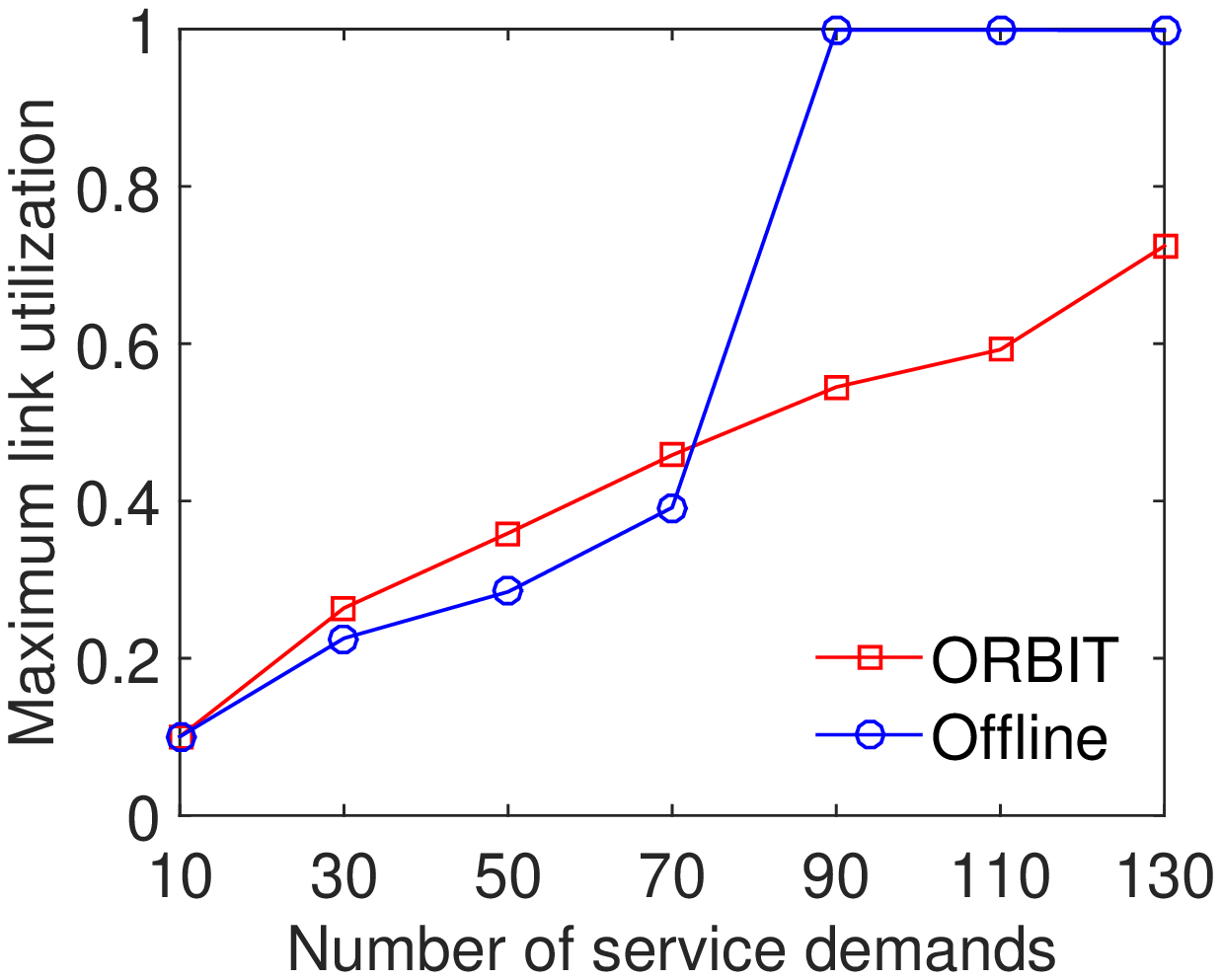}
                }
                ~
                \subfloat[\label{fig-compare-offline-i2-accept}]{
                        \centering
                        \includegraphics[width=0.4\columnwidth]{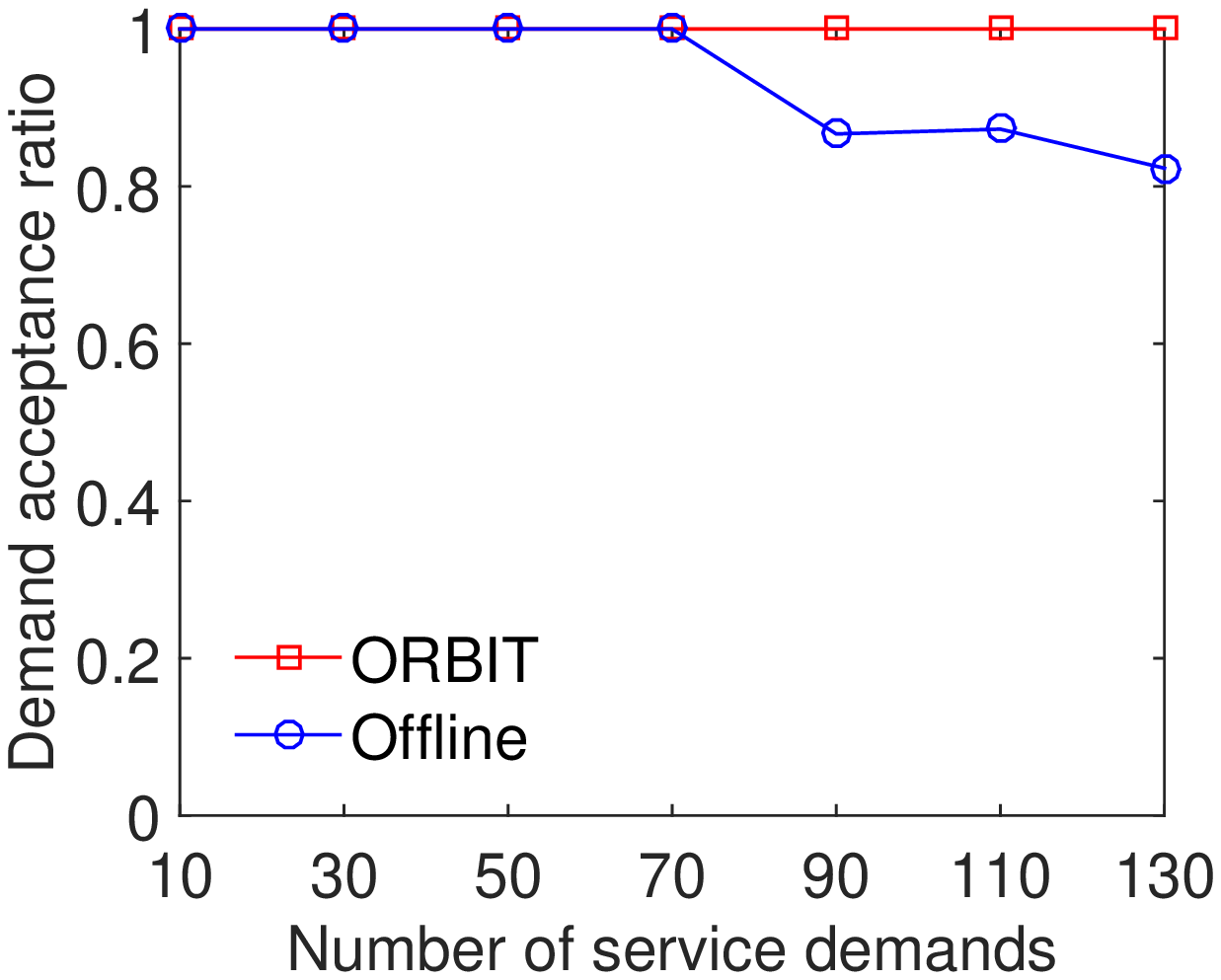}
                }
                \caption{Comparison between ORBIT and an offline solution using Internet2}
                \label{fig-compare-offline-i2}
        \end{figure}

\begin{figure}[!t]
                \centering
                \subfloat[\label{fig-compare-offline-g-link}]{
                        \centering
                        \includegraphics[width=0.4\columnwidth]{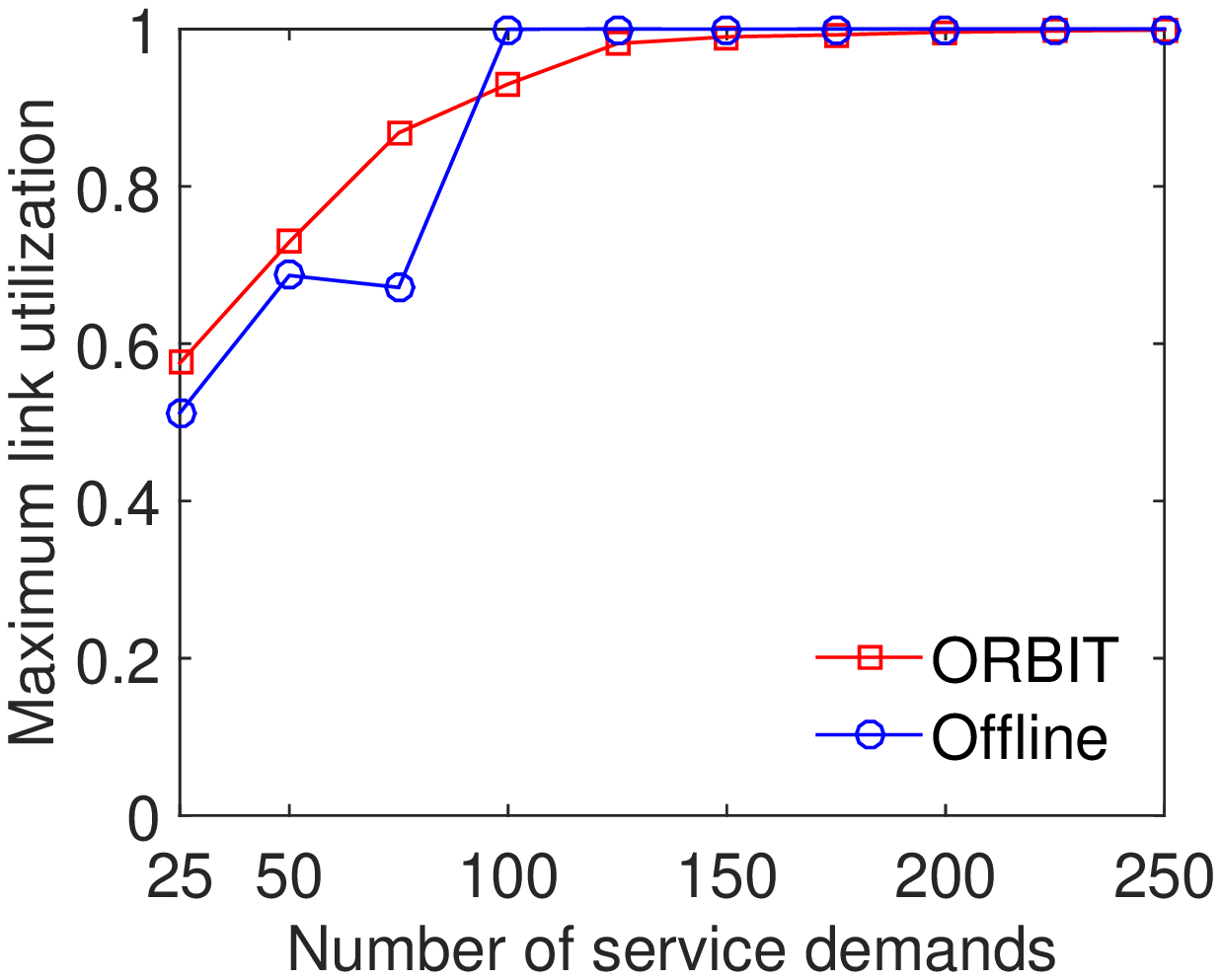}
                }
                ~
                \subfloat[\label{fig-compare-offline-g-accept}]{
                        \centering
                        \includegraphics[width=0.4\columnwidth]{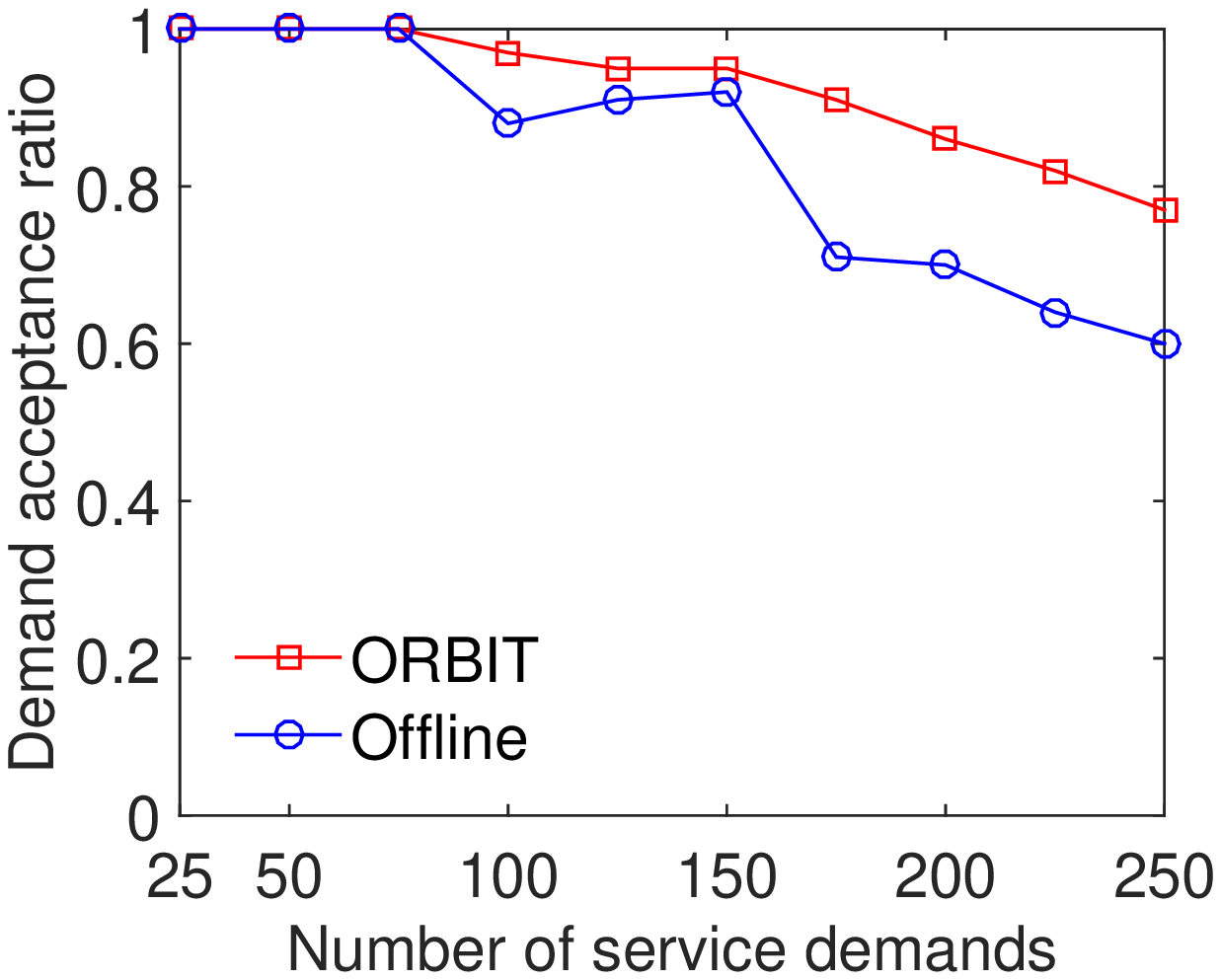}
                }
                \caption{Comparison between ORBIT and an offline solution  using Geant}
                \label{fig-compare-offline-g}
        \end{figure}

        Finally, in order to evaluate the performance of ORBIT in a large scenario we develop an offline solution based on a basic heuristic method called simulated annealing \cite{kirkpatrick1983_optimization} for comparison purposes.
                As shown in Fig. \ref{fig-compare-offline-i2} and \ref{fig-compare-offline-g}, ORBIT outperforms the offline solution in term of both the maximum link utilization and  demand acceptance ratio when the number of service demands is large.
                We argue that ORBIT is more efficient in this case due to the support of near-optimal paths for load balancing in a heavy loaded network.

        The above results demonstrate that we have designed an efficient solution for online load balancing in NFV.
                ORBIT achieves good performance in terms of the link utilization and demand acceptance ratio in comparison with the optimal and offline solutions, especially when the network load is high.

\section{Conclusion}
        \label{s-conclusion}

        We addressed the optimization problem of load balancing in NFV, which is important for guaranteeing the high performance requirements of a network function in a virtualized environment.
                We formulated the problem as a MILP model to compute the optimal solution.
                We developed the ORBIT algorithm that provides an efficient solution for online load balancing.
                As demonstrated through theoretical analyses, the performance of ORBIT is $O\pparen{\varepsilon \log \kappa}$-competitive where $\varepsilon$ and $\kappa$ are the algorithm parameters associated to graph partitioning on NFVI.
                Using two real-world datasets, we performed several experiments in which the result provided by ORBIT is close to the optimal solution.
                Importantly, the experiment results show that ORBIT works very well in comparison with an offline solution when the network load is high.
                This is obviously an important aspect for the practical deployment of our load balancing solution.

        Although focusing on load balancing across multipaths according to ECMP, we believe the model and algorithm developed in this study can be exploited in the procedures of optimizing several performance metrics of other multipath routing schemes in NFV.
                Possible extensions of our results include an evaluation of the impact of ORBIT on congestion control when considering a delay guarantee in a demand,
                a more detailed analysis taking into account demand statistics of end-users in the network partitioning phase of ORBIT,
                or the capability of meeting a Service Level Agreement (SLA) when an infrastructure failure event occurs in NFV.
                It will be valuable to study also the performance impacts and economics of NFV in a mobile and multiple providers environment as in \cite{pham2012_dtn,pham2015_analysis}.

\section*{Acknowledgment}
	This work was partially supported by project B2016-SPH-17 from the Vietnam Ministry of Education and Training.

\bibliographystyle{IEEEtran}
\bibliography{IEEEabrv,pham2017online_techrep}

\end{document}